\documentclass[
 reprint,
superscriptaddress,
superscript,
twocolumn,
showpacs,
footinbib
 amsmath,amssymb,
 aps,
pra,
]{revtex4-1}

\usepackage{epsf}
\usepackage{epsfig}
\usepackage{braket}
\usepackage{amsfonts}
\usepackage{amsmath}
\usepackage{amssymb}
\usepackage{color,soul}
\usepackage{wasysym}
\usepackage{mathrsfs}
\usepackage{multirow}
\usepackage{graphicx}
\usepackage{epstopdf}
\usepackage{dcolumn}
\usepackage{bm}
\usepackage{float}
\usepackage{natbib}
\usepackage{subfigure}
\usepackage[colorlinks=true,
            linkcolor=blue,
            urlcolor=blue,
            citecolor=red,
            filecolor=blue,
            bookmarksopen=true,
            pdfauthor={Mario Miranda Rojas},
            pdftitle={On the control of interference and diffraction of a 3-level atom in a double-slit scheme with cavity fields},
            pdfsubject={On the control of interference and diffraction of a 3-level atom in a double-slit scheme with cavity fields_Manuscript},
            pdfpagemode=UseOutlines]{hyperref}
\usepackage[mathlines]{lineno}

\begin{document}

\preprint{APS/123-QED}

\title{On the control of interference and diffraction of a 3-level atom in a double-slit scheme with cavity fields}
\author{Mario Miranda}
\email{memiranda1@uc.cl}
\affiliation{Instituto de F\'isica, Pontificia Universidad Cat\'olica de Chile, Casilla 306, Santiago, Chile}
\author{Miguel Orszag}
\email{morszag@fis.puc.cl}
\affiliation{Instituto de F\'isica, Pontificia Universidad Cat\'olica de Chile, Casilla 306, Santiago, Chile}
\affiliation{Centro de \'Optica e Informaci\'on Cu\'antica, Universidad Mayor, Camino La Pir\'amide 5750, Huechuraba, Santiago, Chile}

\date{\today}

\begin{abstract}
A double cavity with a quantum mechanical and a classical field is located immediately behind of a double-slit in order to analyse the wave-particle duality. Both fields have common nodes and antinodes through which a three-level atom passes after crossing the double-slit. The atom-field interaction is maximum when the atom crosses a common antinode and path-information can be recorded on the phase of the quantum field. On other hand, if the atom crosses a common node, the interaction is null and no path-information is stored. A quadrature measurement on the quantum field can reveal the path followed by the atom, depending on its initial amplitude $\alpha$ and the classical amplitude $\varepsilon$. In this report we show that the classical radiation acts like a focusing element of the interference and diffraction patterns and how it alters the visibility and distinguishabilily. Furthermore, in our double-slit scheme the two possible paths are correlated with the internal atomic states, which allows us to study the relationship between concurrence and wave-particle duality considering different cases.
\end{abstract}

\maketitle
\section{Introduction}
The Bohr's Principle of Complementarity ~\cite{bohr1928quantum} states that two complementary properties of a given quantum system cannot be obtained simultaneously. This implies that in a measurement process of two complementary observables of a quantum-mechanical object, the total knowledge of the first one makes that all possible outcomes of the second one are equally probable. The wave-particle duality of nature represents the best example of mutually exclusive properties of quantum systems, and several experimental and theoretical works have been developed in order to study this behaviour ~\cite{PhysRev.47.777, PhysRevD.19.473, Aharonov875}. For instance, in a double-slit Young-type scheme, the particle-like  properties are attributed to the knowledge of the path followed by the particle, i.e to the distinguishability ($D$). On other hand, the wave-like properties are associated to the fringe visibility ($V$) on the screen. 

In a double-slit scheme, the obtaining of path-information can be achieved using an external device which acts like a which-path detector ~\cite{scully1997quantum, scully1991quantum}. For instance, if an atom passes through the slits, a quantum field can be located immediately after them and store path-information \cite{PhysRevLett.68.472, PhysRevA.47.405}. This is because the atom-field interaction affects the initial phase of the quantum field depending on the atom's position with respect to the nodes and antinodes of the wave. Thus, if path-information is recorded on the field, it can be extracted by performing a proper measurement in order to know the path followed by the atom and obtain the particle-like properties of the system. However, the stored path-information can also be erased \cite{scully1997quantum,PhysRevA.25.2208,PhysRevA.47.405} in order to restore the wave-like behaviour of the system and thus observing the typical interference pattern on the screen.

In the wave-particle duality the wave-like and the particle-like properties are determined via path-information or fringe visibility and has been quantified mathematically through the inequality  
\begin{equation}\label{englert}
V^{2}+D^{2}\leq 1,
\end{equation}
which has been demonstrated by Englert \cite{PhysRevLett.77.2154} and also derived in other ways \cite{GREENBERGER1988391, PhysRevA.51.54}. Several works have shown that depending on the initial setup of a double-slit experiment, the wave-particle duality can be controlled in order to analyse the complementarity between distinguishability and visibility \cite{PhysRevA.48.1023, JAKOB2010827, orszag2020particle}. Furthermore, it is possible to establish correlations between an intrinsic degree of freedom of the particle passing through the double-slit and the possible paths of the scheme. This implies that the inequality which controls the complementarity between particle and wave, must be modified  as to include this correlation as a third parameter. Recently, concurrence has been considered in a double-slit experiment with single-photons, in order to quantify the established correlations between the paths of the double-slit and the polarization of the photons \cite{JAKOB2010827, PhysRevA.76.052107, PhysRevLett.80.2245, PhysRevResearch.2.012016}. The results have demonstrated that the inequality (\ref{englert}) in presence of the concurrence turns into the equality:
\begin{equation}\label{qian}
V^{2}+D^{2}+C^{2}=1,
\end{equation}
where $C$ represents the degree of quantum entanglement between the polarization of photons and the possible paths of the scheme. Therefore, as a result of the new equality, the definitions of distinguishability and visibility may simultaneously vanish depending on the degree of correlation present in the scheme.  

In this report, instead of photons, we have three level atoms passing through a double slit scheme and immediately after, crossing two cavity fields, one classical (CF) and other quantum mechanical (QF) \cite{Orszag_1995}. Henceforth, we consider $V_{0}$, $D_{0}$ and $C_{0}$ as the respective visibility, distinguishability and concurrence without the cavity fields.  We show that the quantum field acts as a control on the balance between distinguishability and visibility, even to the extreme of reversing their behavior by varying the amount of which-path information coming from the atomic dependent phase of the field, after the interaction and homodyne measurement. On the other hand, the  classical field produces a ``focusing effect" in the sense that for larger field, the interference plane becomes closer to the slit-cavity setup, so it can be used  to control the path information stored in the quantum field and modify the pattern observed on the screen.

\section{Model}
In this article we consider a three-level atom crossing a double cavity with a quantum and a classical field (Fig.~\ref{Figure3}). The fields have wave numbers $k=2\pi/\lambda_{QF}=3k'$ and $k'=2\pi/\lambda_{CF}$ respectively. A double-slit is located immediately before the fields, with the top slit in front of a common antinode and the bottom slit in front of a common node. The separation distance between slits is $0.75\lambda_{QF}=0.25\lambda_{CF}$.

\begin{figure}[h!]\centering
\includegraphics[width=8.6cm]{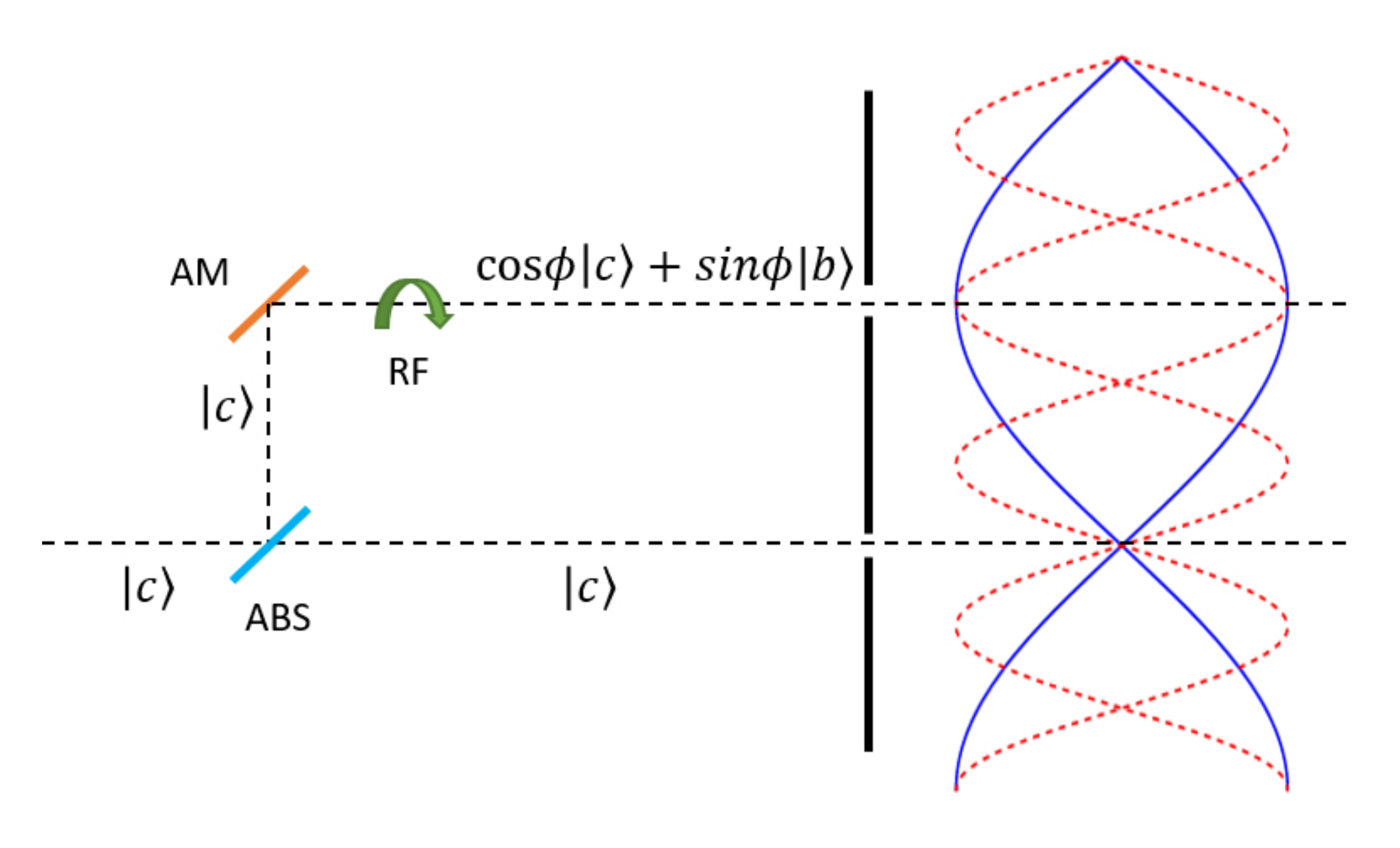}\\
\caption{\label{Figure1}Scheme of the possible paths followed by the atom. ABS: Atomic Beam Splitter, AM: Atomic Mirror, RF: Ramsey Field. The atom is either reflected or transmitted by the ABS by taking the upper or lower path, respectively. Finally, the atom crosses the double-slit and both, quantum (red) and classical (blue) fields.}
\end{figure}

Previous to the double slit, the spatial atomic state is realized by an atomic beam splitter (ABS) \cite{PhysRevA.43.2455, PhysRevLett.56.827} and an atomic mirror (AM) \cite{PhysRevLett.60.2137, Merimeche_2006}, and the internal atomic state in the top path is realized by a Ramsey field (RF) \cite{PhysRev.78.695} (Fig.~\ref{Figure1}). The reflection and transmission coefficients of the ABS are $c_{\uparrow}$ and $c_{\downarrow}$, satisfying $|c_{\uparrow}|^{2}+|c_{\downarrow}|^{2}=1$. If the atom is transmitted, it flies along the bottom path and crosses the slit at the node of the standing waves in the position $x=0.75\lambda_{QF}$. On other hand, if the atom is reflected, it goes through the top slit using a AM and then a RF. The task of the RF is to prepare a superposition of the ground state $|c\rangle$ and the intermediate state $|b\rangle$. Here the probability coefficients of exciting the state $|b\rangle$ and remaining in the state $|c\rangle$ are $\sin^{2}\phi$ and $\cos^{2}\phi$, respectively. In this case, the atom crosses the top slit and passes through the common antinode of the fields in the position $x=0$. Therefore, the top path is correlated with the internal atomic state $|\Phi_{\uparrow}\rangle=\cos\phi|c\rangle+\sin\phi|b\rangle$, while the bottom path is correlated with the state $|\Phi_{\downarrow}\rangle=|c\rangle$ .
\subsection{Initial state}
Initially the atom is in the ground state $|c\rangle$. After passing through the ABS and considering the effect of the AM and the RF, the atomic state can be described as
\begin{equation}
\begin{split}
|\psi(0)\rangle_{atom}&=c_{\uparrow}|P_{\uparrow}\rangle\otimes|\Phi_{\uparrow}\rangle+c_{\downarrow}|P_{\downarrow}\rangle\otimes|\Phi_{\downarrow}\rangle\\
&=c_{\uparrow}|P_{\uparrow}\rangle\otimes[\cos\phi|c\rangle+\sin\phi|b\rangle]+c_{\downarrow}|P_{\downarrow}\rangle\otimes|c\rangle,
\end{split}
\end{equation}
where the states $|P_{\uparrow}\rangle$ and $|P_{\downarrow}\rangle$ represent the top and bottom path of the scheme, respectively.

\begin{figure}[h!]\centering
\includegraphics[width=6cm]{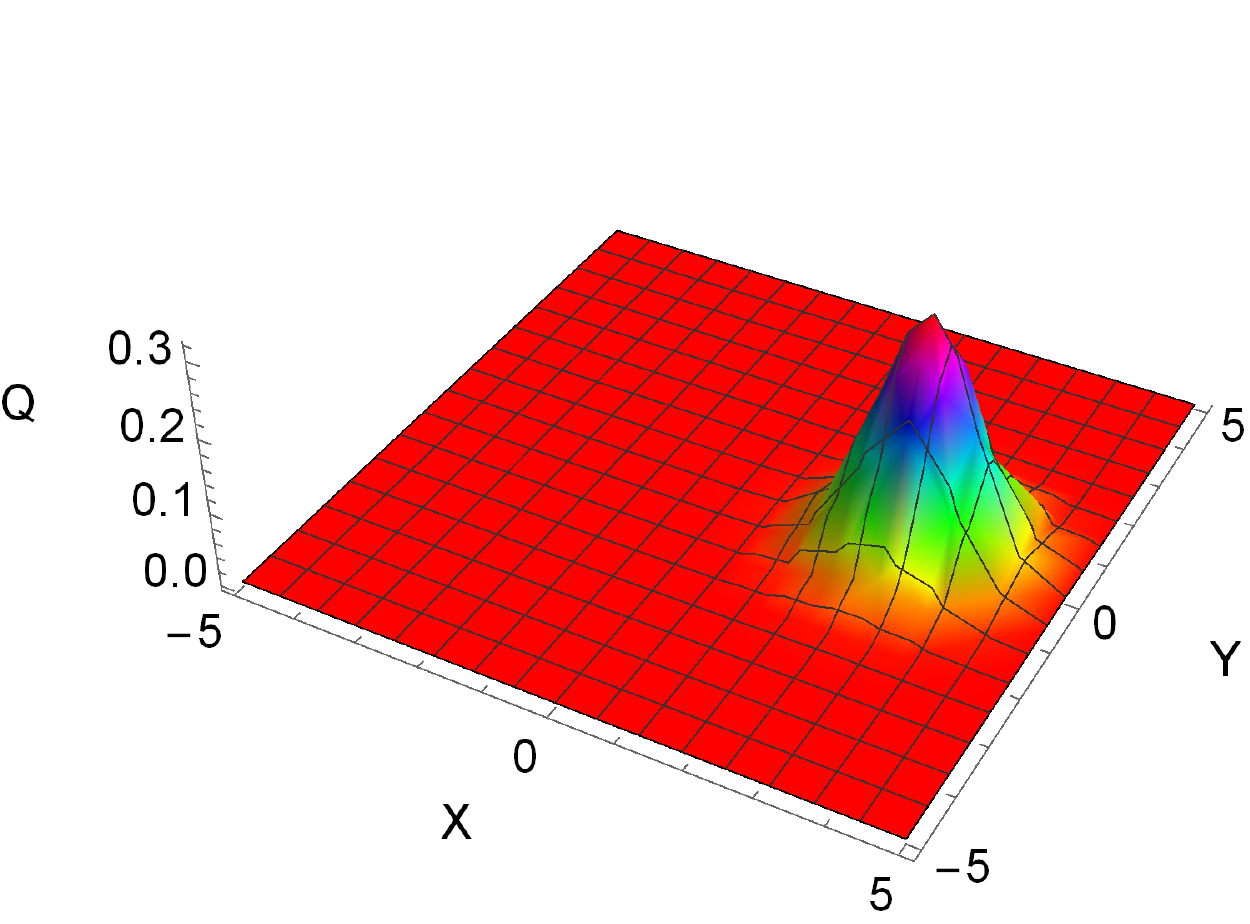}\\
\caption{\label{Figure2}Initial phase of the quantum field $|\alpha\rangle$ for an amplitude $\alpha=\sqrt{8}$, where $X$ and $Y$ correspond to the amplitude and phase quadrature of the field, respectively.}
\end{figure}

Immediately to the right of the double slit, a double cavity with both, classical and quantum fields is located. The quantum field before the interaction is a coherent state with amplitude $\alpha=\sqrt{8}$ (Fig.~\ref{Figure2}),
\begin{equation}
|\psi(0)\rangle_{field}=|\alpha\rangle = e^{\frac{-|\alpha|^{2}}{2}}\sum_{m=0}\frac{\alpha^{m}}{\sqrt{m!}}|m\rangle=\sum_{m=0}c_{m}|m\rangle,
\end{equation}
and the total initial system is given as
\begin{equation}
\begin{split}
&|\psi(0)\rangle_{system} =|\psi(0)\rangle_{atom}\otimes|\psi(0)\rangle_{field}\\
&=\bigg( c_{\uparrow}|P_{\uparrow}\rangle\otimes[\cos\phi|c\rangle+\sin\phi|b\rangle]+c_{\downarrow}|P_{\downarrow}\rangle\otimes|c\rangle \bigg) \otimes |\alpha\rangle.\\
\end{split}
\end{equation}

\subsection{Time evolution of the system}
\begin{figure}[h!]\centering
\subfigure[\label{Figure3a}Three-level atom.]{\includegraphics[width=4cm]{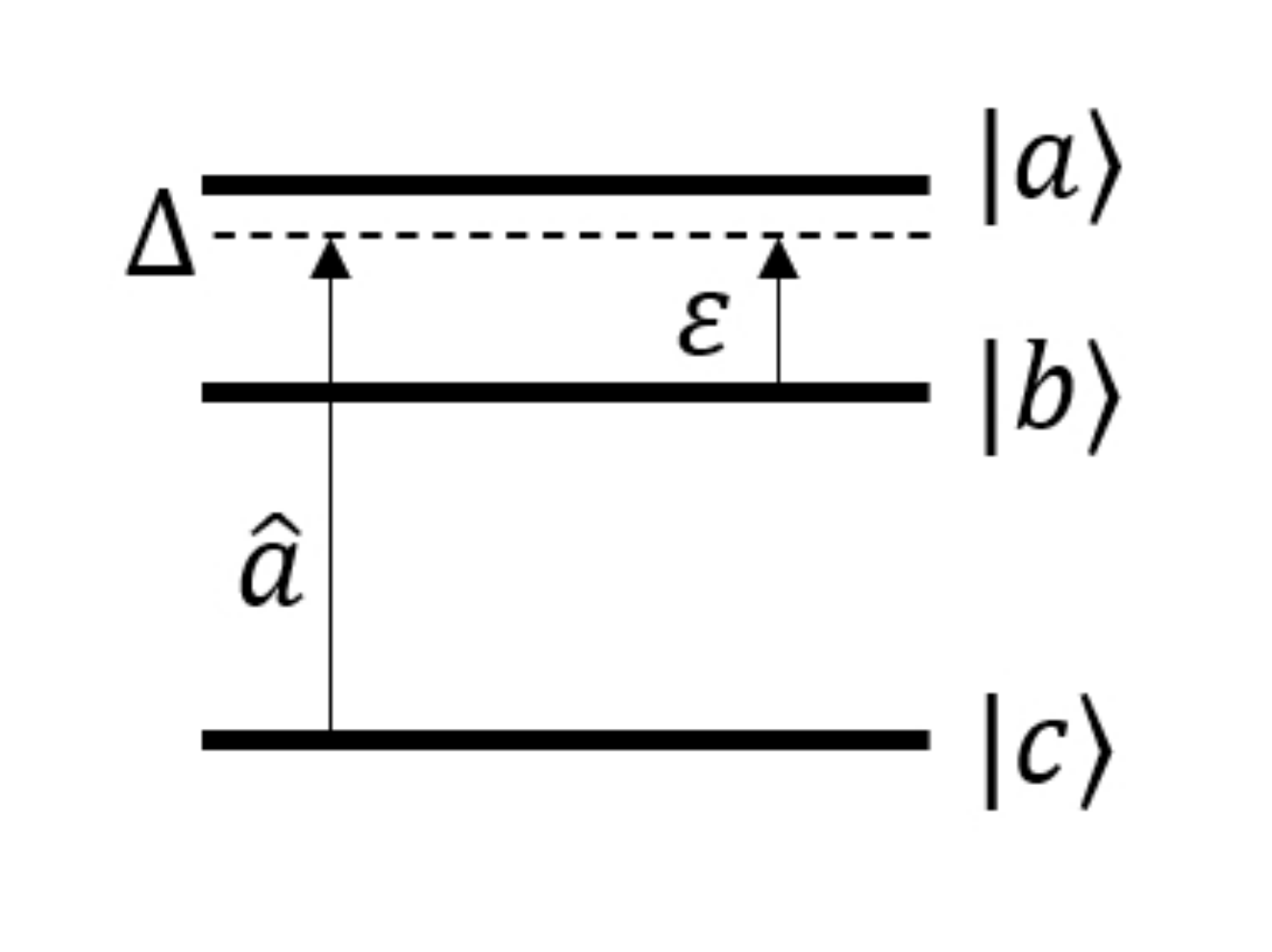}}
\subfigure[\label{Figure3b}Double cavity.]{\includegraphics[width=7cm]{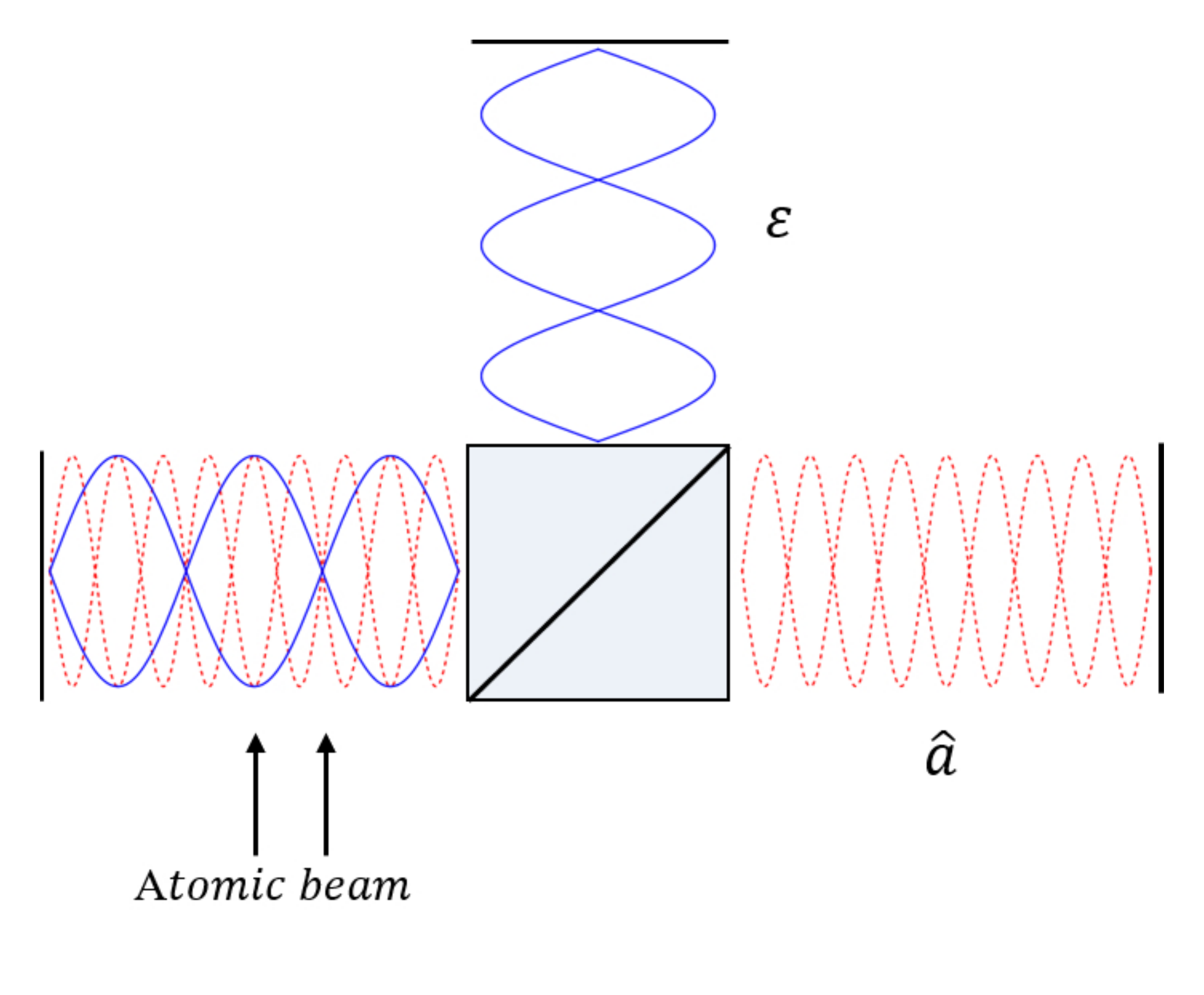}}
\caption{\label{Figure3}A three-level atom crosses the double cavity with a quantum (red) and a classical (blue) field.} 
\end{figure}

After the interaction the total initial system will evolve to
\begin{equation}
|\psi(t)\rangle_{system}=\hat{U}|\psi(0)\rangle_{system}=e^{-\frac{i\hat{V}t}{\hbar}}|\psi(0)\rangle_{system},
\end{equation}
where $\hat{V}$ is the Hamiltonian in the interaction framework considering a rotating wave approximation,
\begin{equation}
\begin{split}
\hat{V}&=\hbar g_{1} \big( \hat{a} e^{i\Delta t}|a\rangle\langle c|+ \hat{a}^{\dag} e^{-i\Delta t}|c\rangle\langle a|\big)\\
&+\hbar g_{2} \big( \varepsilon e^{i\Delta t}|a\rangle\langle b|+ \varepsilon^{*} e^{-i\Delta t}|b\rangle\langle a|\big).
\end{split}
\end{equation}
Here the quantum field $\hat{a}$ couples the $|a\rangle - |c\rangle$ transition, while the classical field $\varepsilon$ couples the $|a\rangle - |b\rangle$ transition with coupling constant $g_{1}=g \cos(kx)$ and $g_{2}=g' \cos(k'x)$ respectively, where $k'=k/3$. For both fields, the detuning $\Delta$ is the same and it is required to be large in order to avoid photon emission and therefore, an effect on the cavity field (Fig.~\ref{Figure3}).
The elements of the evolution operator for this system are given by \cite{Orszag_1995}
\begin{equation}
\begin{split}
&\bullet U_{aa}= e^{i\Delta t/2}\bigg( R - \frac{i\Delta}{2}S\bigg)\quad \bullet U_{ab}=-ig_{2}\varepsilon e^{i\Delta t/2}S\\
&\bullet U_{ac}=-ig_{1} e^{i\Delta t/2}S\hat{a} \quad \bullet U_{ba}=-ig_{2}\varepsilon^{*} e^{-i\Delta t/2}S\\
&\bullet U_{bb}=1+\frac{g_{2}^{2}|\varepsilon|^{2}\big[e^{-i\Delta t/2 } \big( R+\frac{i\Delta}{2}S\big)-1\big]}{\Lambda} \\
&\bullet U_{bc}=g_{1}g_{2}\varepsilon^{*}\frac{\big[e^{-i\Delta t/2 } \big( R+\frac{i\Delta}{2}S\big)-1\big]\hat{a}}{\Lambda}\\
&\bullet U_{ca}=-ig_{1}\hat{a}^{\dag}e^{-i\Delta t/2}S\\
&\bullet U_{cb}=g_{1}g_{2}\varepsilon \hat{a}^{\dag}\frac{\big[ e^{-i\Delta t/2} \big( R + \frac{i\Delta}{2}S\big) -1\big]}{\Lambda}\\
&\bullet U_{cc}=1+\frac{g_{1}^{2}\hat{a}^{\dag}\hat{a}\big[e^{-i\Delta t/2 } \big( \overline{R}+\frac{i\Delta}{2}\overline{S}\big)-1\big]}{\overline{\Lambda}}
\end{split}
\end{equation}
where
\begin{equation}
\begin{split}
&\Lambda \equiv g_{2}^{2} |\varepsilon|^{2} +  g_{1}^{2}\hat{a}\hat{a}^{\dag},\quad \overline{\Lambda}\equiv g_{2}^{2} |\varepsilon|^{2} +  g_{1}^{2}\hat{a}^{\dag}\hat{a}, \quad S\equiv \frac{\sin\sqrt{\mu}t}{\sqrt{\mu}},\\
& \overline{S} \equiv \frac{\sin\sqrt{\overline{\mu}}t}{\sqrt{\overline{\mu}}}, \quad R \equiv \cos\sqrt{\mu}t, \quad \overline{R} \equiv \cos\sqrt{\overline{\mu}}t,\\
&\mu \equiv g_{2}^{2} |\varepsilon|^{2} + g_{1}^{2}\hat{a}\hat{a}^{\dag}+\Delta^{2}/4, \quad \overline{\mu} \equiv g_{2}^{2} |\varepsilon|^{2} + g_{1}^{2}\hat{a}^{\dag}\hat{a}+\Delta^{2}/4.
\end{split}
\end{equation}
For arbitrary paths, the state of the system after a time of interaction $t$ can be written as 
\begin{equation}\label{evolution}
\begin{split}
|\psi(t)\rangle&=c_{\uparrow}\cos\phi|P_{\uparrow}\rangle\otimes \hat{U}|c\rangle \otimes |\alpha\rangle+c_{\uparrow}\sin\phi|P_{\uparrow}\rangle\otimes \hat{U}|b\rangle \otimes |\alpha\rangle\\
&+ c_{\downarrow}|P_{\downarrow}\rangle\otimes \hat{U}|c\rangle  \otimes |\alpha\rangle\\
&=c_{\uparrow}\cos\phi|P_{\uparrow}\rangle\otimes \big[U_{bc}|b\rangle+U_{cc}|c\rangle\big] \otimes |\alpha\rangle\\
&+c_{\uparrow}\sin\phi|P_{\uparrow}\rangle\otimes \big[U_{bb}|b\rangle+U_{cb}|c\rangle\big] \otimes |\alpha\rangle \\
&+ c_{\downarrow}|P_{\downarrow}\rangle\otimes \big[U_{bc}|b\rangle+U_{cc}|c\rangle\big] \otimes |\alpha\rangle\\
&=c_{\uparrow}\cos\phi|P_{\uparrow}\rangle\otimes \bigg[\sum \beta_{m}^{c}|m-1\rangle|b\rangle+\sum \alpha_{m}^{c}|m\rangle|c\rangle\bigg]\\
&+c_{\uparrow}\sin\phi|P_{\uparrow}\rangle\otimes \bigg[\sum \alpha_{m}^{b}|m\rangle|b\rangle+\sum \beta_{m}^{b}|m+1\rangle|c\rangle\bigg] \\
&+ c_{\downarrow}|P_{\downarrow}\rangle\otimes \bigg[\sum \beta_{m}^{c}|m-1\rangle|b\rangle+\sum \alpha_{m}^{c}|m\rangle|c\rangle\bigg],\\
\end{split}
\end{equation}
where the coefficients $\alpha_{m}^{b,c}$ and $\beta_{m}^{b,c}$ depend on the internal state of atom (see Appendix \ref{appa}).

\subsection{Quadrature measurement}
In this model the which-path information depends on the phase-shift of the quantum field as a consequence of the atom's position during the interaction time $t$. As mentioned before, the maximum atom-field interaction is accomplished when the atom takes the top path and crosses the common antinode of both fields. In that case, we must consider the two possible internal states of the atom, $|b\rangle$ and $|c\rangle$, and the effect of these on the quantum field \cite{PhysRevA.47.405}. On other hand, if the atom passes through the bottom slit and then crosses the common node, no interaction occurs, and the initial phase of the field remains the same (see \ref{nointeraction} in Appendix \ref{appa}). Therefore, considering the phase-shift caused either by the ground or intermediate atomic state in the top path, a quadrature measurement could reveal the path followed by the atom.

If the atom crosses the common antinode ($c_{\uparrow}=1$) in the state $|b\rangle$ ($\phi=\pi/2$) or $|c\rangle$ ($\phi=0$), the final state of the total system after interaction corresponds to a superposition of the internal states $|b\rangle$ and $|c\rangle$ given respectively by
\begin{equation}\label{statebi}
|\psi(t)\rangle_{system}^{b}=|P_{\uparrow}\rangle\otimes\bigg[\sum_{m}\alpha^{b}_{m}|m\rangle|b\rangle+\sum_{m}\beta^{b}_{m}|m+1\rangle|c\rangle\bigg],
\end{equation}
and
\begin{equation}\label{stateci}
|\psi(t)\rangle_{system}^{c}=|P_{\uparrow}\rangle\otimes\bigg[\sum_{m}\alpha^{c}_{m}|m\rangle|c\rangle+\sum_{m}\beta^{c}_{m}|m-1\rangle|b\rangle\bigg].
\end{equation}
Therefore, considering the effect of both, quantum and classical fields on the internal atomic state, the evolution of the total system can be understood as a Raman diffraction process in which the internal atomic state is changed, or a Bragg diffraction process where the internal state of the atom remains unaffected \cite{PhysRevA.101.053610, abend2020atom}. These processes can be controlled by the amplitude of the classical field, since that for small values of $\varepsilon$ the coefficients $\beta^{b,c}_{m}$ decrease and it is more probable that the atom remains in its initial state, while as $\varepsilon$ increases, the transition from $|b\rangle$ to $|c\rangle$ or vice versa becomes more probable.
For simplicity, we first consider only the quantum field in order to analyse the effects of the atomic state on it. For the specific values of the parameters $\varepsilon=0$, $\alpha=\sqrt{8}$, $g=g'$ and $|g|^{2}t/\Delta=\pi$, equations (\ref{statebi}) and (\ref{stateci}) can be written as
\begin{equation}\label{state b}
\begin{split}
|\psi(t)\rangle_{system}^{b}&=|P_{\uparrow}\rangle\otimes\bigg[\sum_{m}e^{-\frac{|\alpha|^{2}}{2}}\frac{\alpha^{m}}{\sqrt{m!}}|m\rangle|b\rangle\bigg]\\
&= |P_{\uparrow}\rangle\otimes|\alpha\rangle\otimes|b\rangle,
\end{split}
\end{equation}
and
\begin{equation}\label{state c}
\begin{split}
|\psi(t)\rangle_{system}^{c}&=|P_{\uparrow}\rangle\otimes\bigg[\sum_{m}e^{i\pi\cos^{2}(kx)m}e^{-\frac{|\alpha|^{2}}{2}}\frac{\alpha^{m}}{\sqrt{m!}}|m\rangle|c\rangle\bigg]\\
&=|P_{\uparrow}\rangle\otimes|e^{i\eta(x)}\alpha\rangle\otimes|c\rangle,
\end{split}
\end{equation}
respectively, with $\eta(x)=\pi\cos^{2}(kx)$. 

\begin{figure}[h!]\centering
\subfigure[\label{Figure4a}Setup corresponding to the case $\phi=\pi/2$, in which the internal atomic state in the upper path is $|b\rangle$.]{\includegraphics[width=7cm]{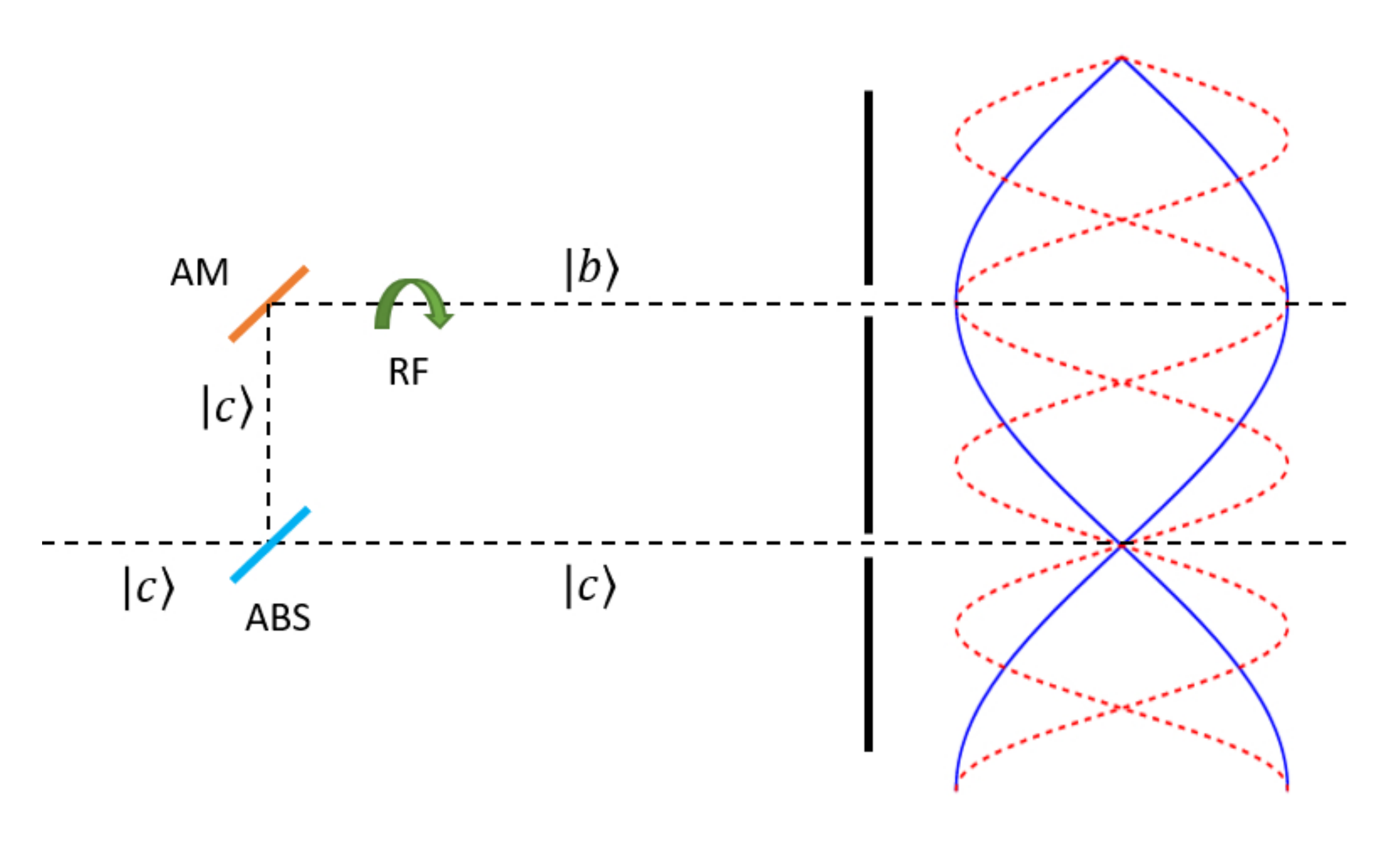}}
\subfigure[\label{Figure4b}Phase evolution after atom-field interaction for $\phi=\pi/2$. The initial phase remains unaffected. The blue plane shows the most probable result ($\chi_{\theta=0}=+\alpha$) if a $X$ quadrature measurement is performed.]{\includegraphics[width=6cm]{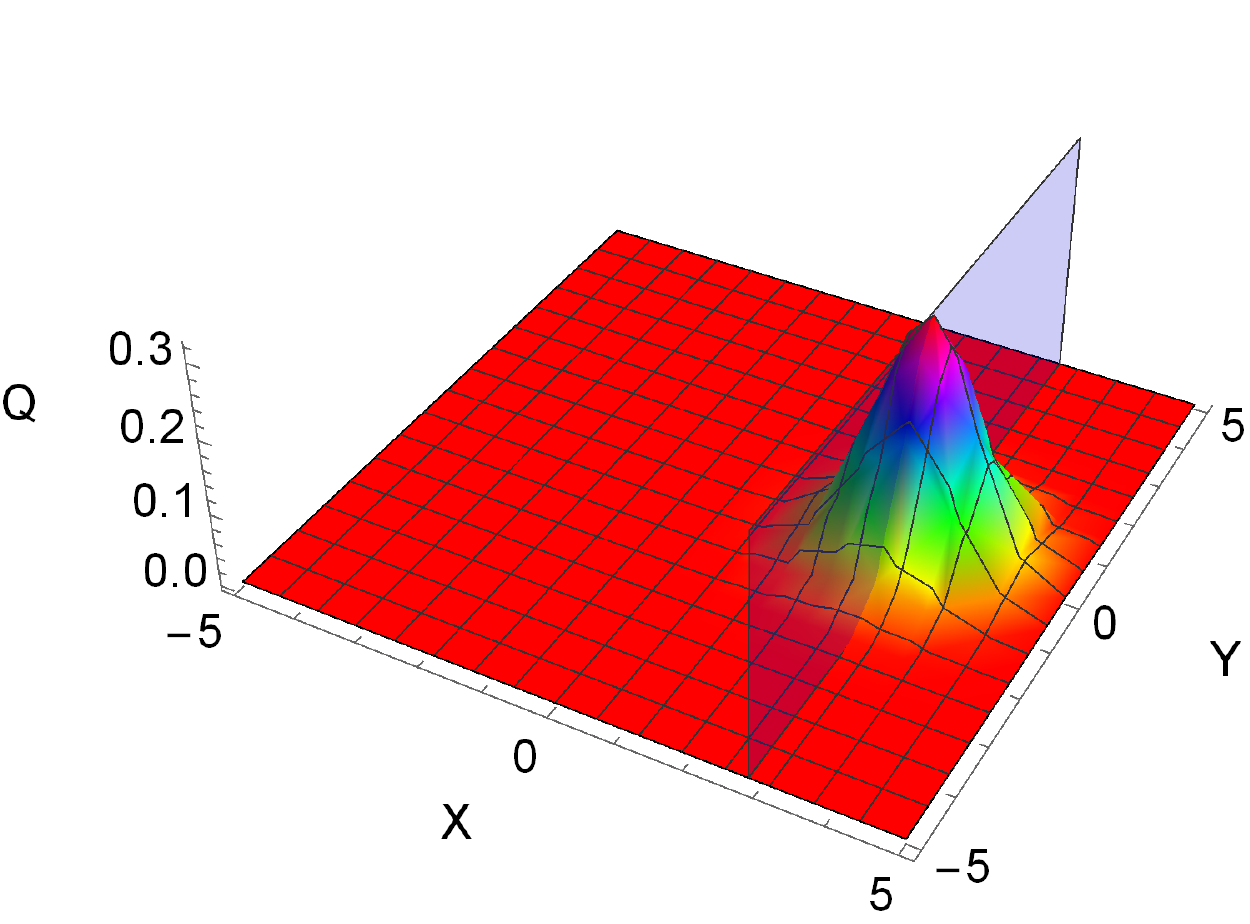}}
\caption{\label{Figure4}If the internal atomic state in the top path is $|b\rangle$, there is no phase-shift in the quantum field for $\alpha=\sqrt{8}$ and $\varepsilon=0$. Therefore, no path-information is record on the field. However, the own internal atomic states in the top and bottom path can give information about which slit the atom passed through.} 
\end{figure}

\begin{figure}[h!]\centering
\subfigure[\label{Figure5a}Setup corresponding to the case $\phi=0$. In this case the internal atomic state in the upper path is $|c\rangle$ and the interaction with the field is maximum.]{\includegraphics[width=7cm]{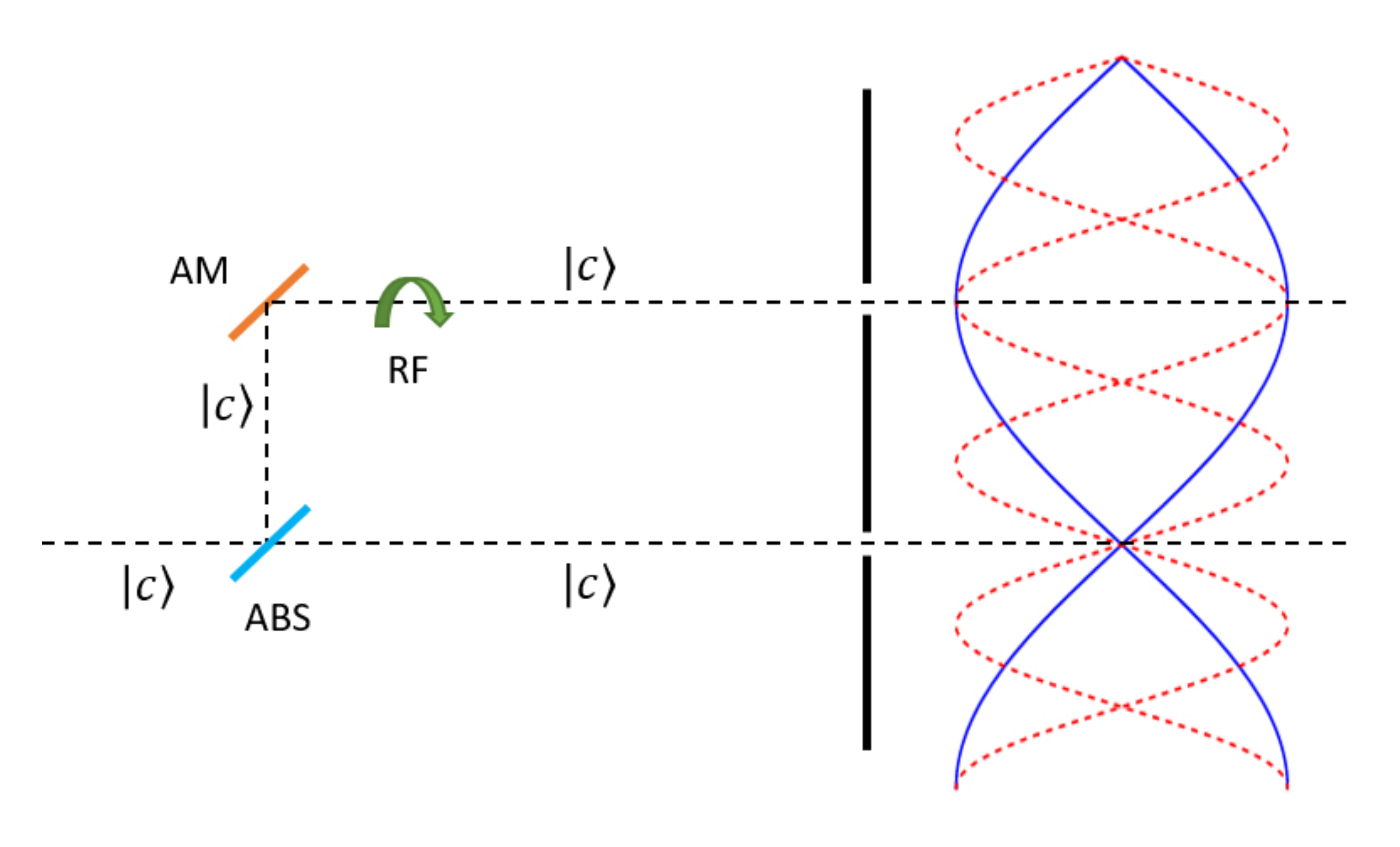}}
\subfigure[\label{Figure5b}Phase evolution after atom-field interaction for $\phi=0$. The initial phase changes from $0$ to $\pi$.]{\includegraphics[width=6cm]{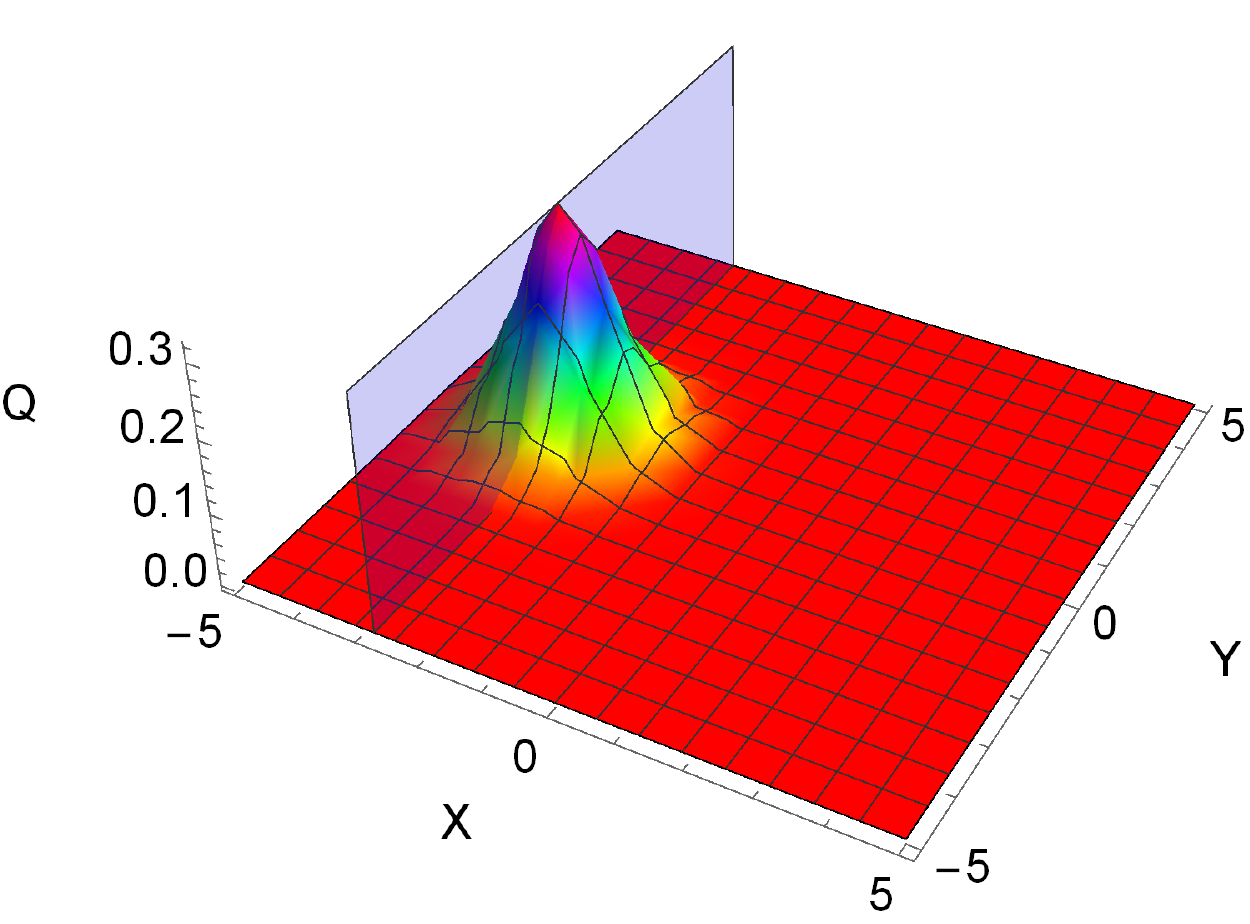}}
\caption{\label{Figure5}If the internal atomic state in the top path is $|c\rangle$, it produces a phase-shift of $\pi$ in the quantum field, which reveals path-information. We consider $\alpha=\sqrt{8}$ and $\varepsilon=0$. In this case the most probable result for a $X$ quadrature measurement is $\chi_{\theta=0}=-\alpha$.} 
\end{figure}

Therefore, when the atom crosses the antinode of the quantum field in the intermediate state $|b\rangle$ [Fig.~\ref{Figure4a}], there is no phase-shift [Fig.~\ref{Figure4b}] and no quadrature measurement can reveal which-path information. This is because the same phase can be obtained if the atom takes the bottom path (initial phase unaffected).
In contrast, when the atom crosses the antinode in the ground state $|c\rangle$ [Fig.~\ref{Figure5a}], the phase increases from $0$ to $\pi$ [Fig.~\ref{Figure5b}]. In that case, the internal atomic state does not reveal path-information by itself. However, the path-information is stored in the phase of the quantum field and can be extracted through a $X$ quadrature measurement.

In general, if the quadrature
\begin{equation}
X_{\theta}=\frac{a e^{-i\theta}+a^{\dag} e^{i\theta}}{2}
\end{equation}
is measured with an eigenvalue $\chi_{\theta}$, the corresponding eigenstate $|\chi_{\theta}\rangle$ is an infinitely squeezed state given by \cite{PhysRevA.47.405, Orszag_1995}
\begin{equation}
\begin{split}
|\chi_{\theta}\rangle &= \frac{1}{\sqrt[4]{2\pi}}\mbox{exp} \big[-\frac{1}{2}(a^{\dag}e^{i\theta}-\chi_{\theta})^{2}+\frac{1}{4}\chi_{\theta}^{2}\big]|0\rangle=\sum_{n}b_{n}|n\rangle,\\
\end{split}
\end{equation}
where
\begin{equation}
b_{n}=\frac{N}{\sqrt{n!}}(\frac{1}{2}e^{i\theta})^{n/2}H_{n}(z),
\end{equation}
with $N$ being a normalization constant. The function $H_{n}(z)$ corresponds to the Hermite polynomials with $z=(\alpha e^{-i\theta}+\alpha^{*} e^{i\theta})/2$.

\begin{figure}[h!]\centering
\subfigure[\label{Figure6a}If the phase remains unaffected, the most probable result for a $Y$ quadrature measurement is $\chi_{\theta=\pi/2}=0$.]{\includegraphics[width=6cm]{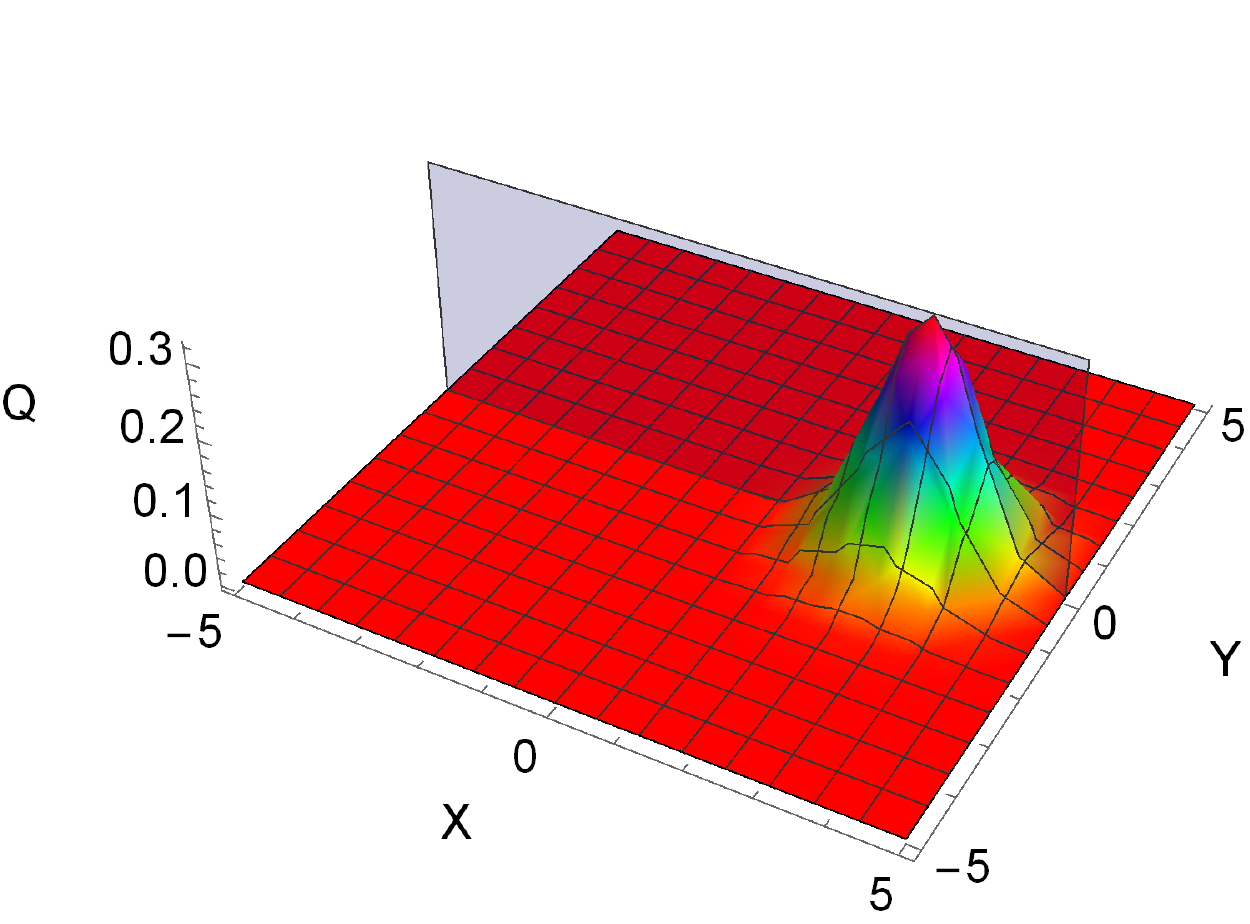}}
\subfigure[\label{Figure6b}If the phase changes from $0$ to $\pi$, the result of a $Y$ quadrature measurement remains the same.]{\includegraphics[width=6cm]{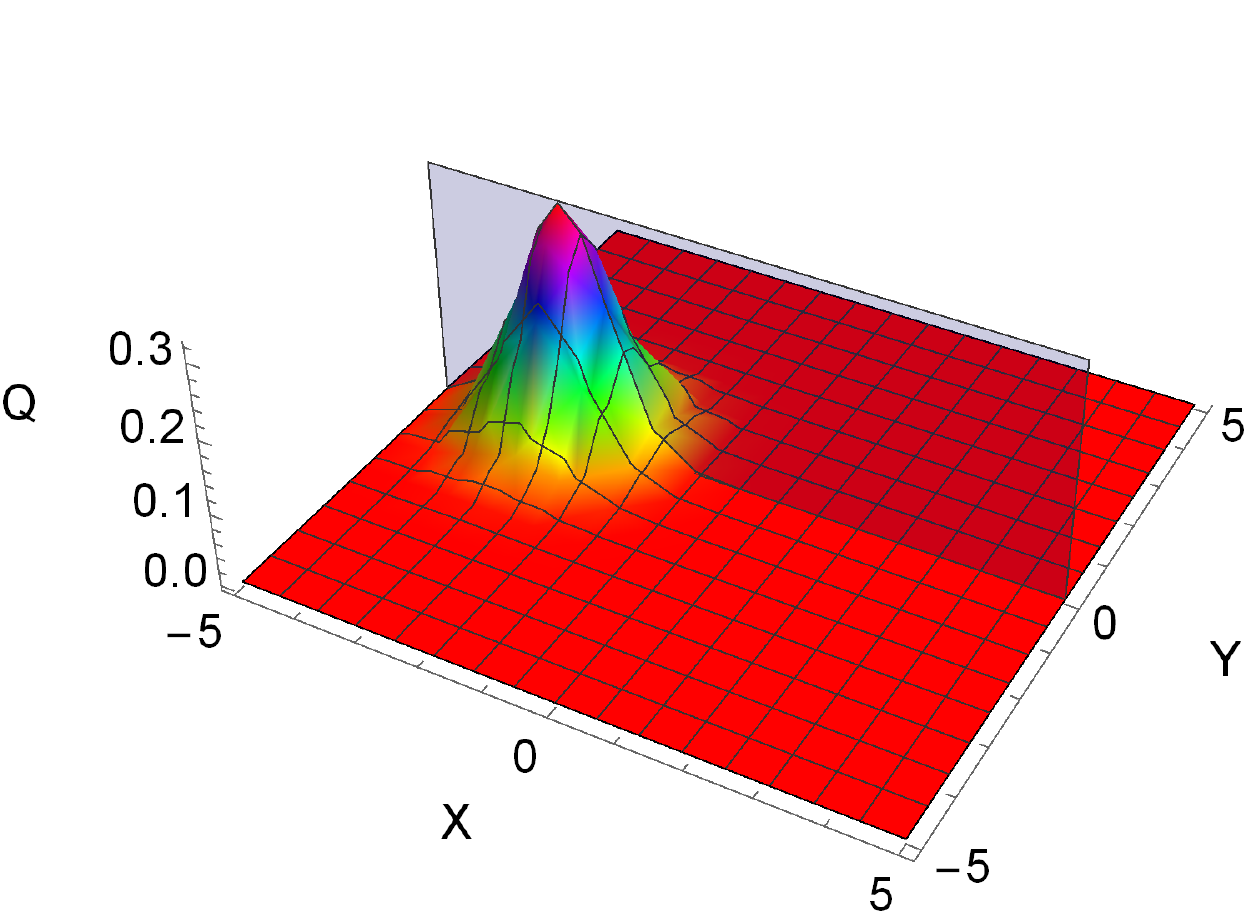}}
\caption{\label{Figure6}A $Y$ quadrature measurement does not reveal path information, because the most probable result is obtained regardless of quantum field state.} 
\end{figure}

Since we consider $|g|^{2}t/\Delta=\pi$, a $X_{\theta=0}=X$ quadrature measurement with values $\chi_{\theta=0}=\pm\alpha$ determines the phase of the field and then we can know whether the atom passed through either the node or antinode (considering $\phi=0$). On other hand, if a $X_{\theta=\pi/2}=Y$ quadrature measurement is performed, and the most probable result is obtained ($\chi_{\theta=\pi/2}=0$), no path-information is obtained and interference appears on the screen, since that from the most probable result no path information is inferred (Fig.~\ref{Figure6}).  

\subsection{Particle-wave duality and concurrence}
In a typical double-slit scheme we can configure several cases in order to study the quantum duality between distinguishabilily (particle-like) and visibility (wave-like) \cite{PhysRevLett.77.2154}. Now, if a correlation is established between some intrinsical property of a particle and the possible paths, the wave-particle duality can be modified depending on the degree of entanglement in the system. Recently, it has been experimentally proven the relation among distinguishabilily ($D_{0}$), visibility ($V_{0}$) and concurrence ($C_{0}$) \cite{PhysRevResearch.2.012016} which can be written as
\begin{equation}\label{dvc}
D_{0}^{2}+V_{0}^{2}+C_{0}^{2}=1,
\end{equation}
with
\begin{equation}
\begin{split}
&D_{0}=||c_{\uparrow}|^{2}-|c_{\downarrow}|^{2}|\\
&V_{0}=2|c_{\uparrow}c_{\downarrow}\gamma|\\
&C_{0}=2|c_{\uparrow}c_{\downarrow}|\sqrt{1-|\gamma|^{2}},
\end{split}
\end{equation}
\cite{GREENBERGER1988391,PhysRevA.51.54,PhysRevA.48.1023, PhysRevD.19.473} where $c_{\uparrow}$ and $c_{\downarrow}$ are coefficients that define the probabilities for the atom of taking the top or bottom path, while $\gamma\equiv\langle \Phi_{\uparrow}|\Phi_{\downarrow}\rangle$, where the normalized states $|\Phi_{\uparrow,\downarrow}\rangle$ correspond to intrinsic degrees of freedom of the particle, in our case the internal atomic state.

\begin{figure}[h!]\centering
\includegraphics[width=6cm]{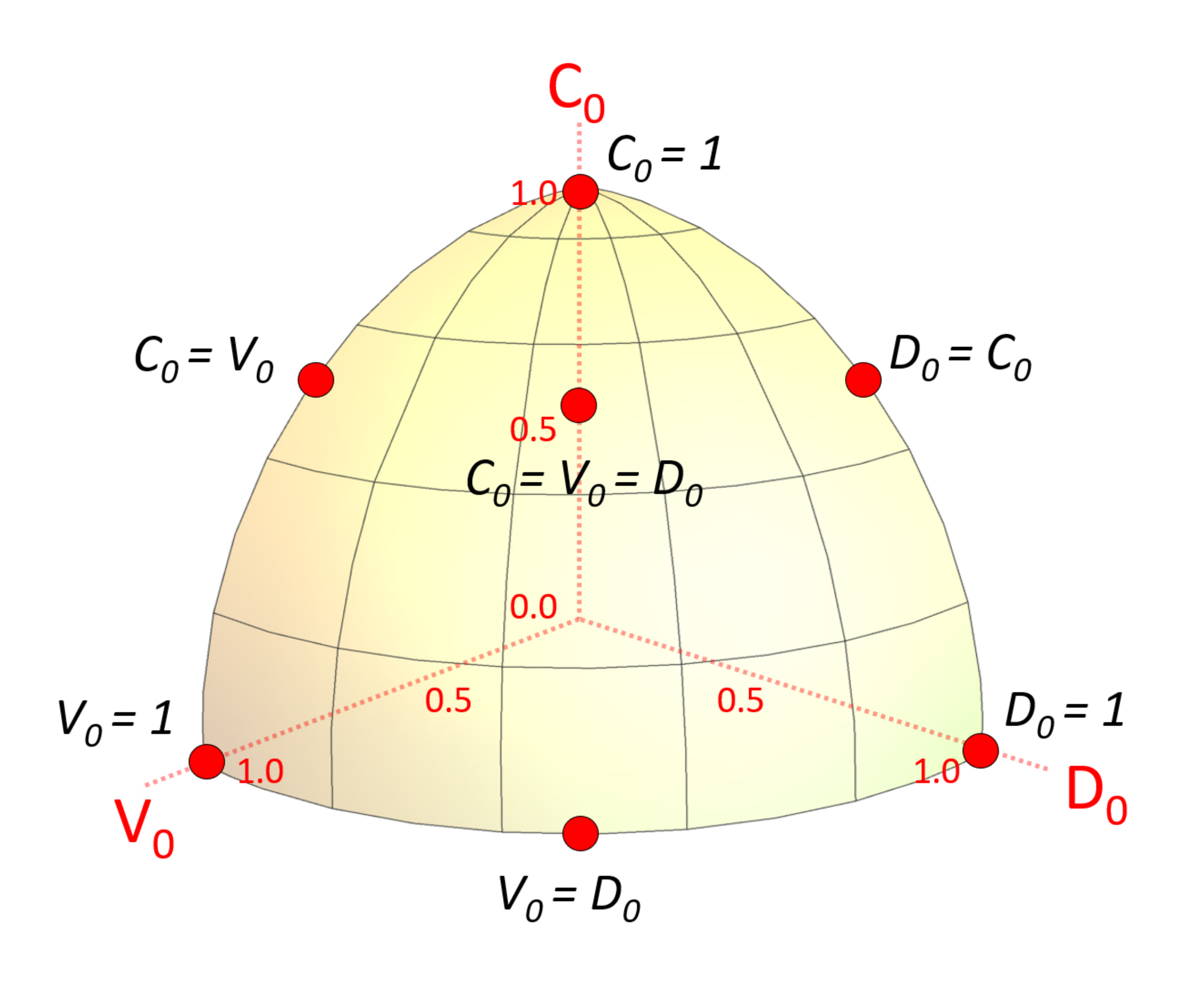}\\
\caption{Unit sphere $D_{0}^{2}+V_{0}^{2}+C_{0}^{2}=1$. The extreme cases $V_{0}=1$, $D_{0}=1$, $C_{0}=1$ and intermediate ones are shown on the surface by red dots.}\label{Figure7}
\end{figure}

Cases of special interest are shown on the surface of the sphere in the Fig.~\ref{Figure7}. The point $C_{0}=1$, with coefficients  $c_{\uparrow}=c_{\downarrow}=1/\sqrt{2}$ and $\gamma=0$, represents a special scenario in which, based on the definitions of $D_{0}$ and $V_{0}$, visibility and distinguishability are equal to zero. So, what would we expect to observe on the screen after the double slit?

In the next section we analyze different cases considering our scheme, in which the which-path information can be stored in the phase-shift of the quantum field, but also it can be controlled through the coefficients $c_{\uparrow}$ and $c_{\downarrow}$, and we show the different patterns that are obtained in each case shown on the sphere. Finally, we show how the classic field can change the initial visibility and which-path information as $\varepsilon$ increases from $0$ to higher values and how the corresponding patterns are modified.

\section{Numerical results}
In the previous sections we explained how the atom can modify the quantum field and how the path-information can be extracted by performing a quadrature measurement. The localization of the atom results in loss of interference and the total knowledge of the path-information. In this section we assume that once the atom leaves the cavity, it freely evolves during a time $t'$ (in units of $2m/\hbar k'^{2}$) to state 
\begin{equation}
\begin{split}
\rho_{atom}(t')&=\hat{U}\rho_{atom}(t)\hat{U}^{\dag}\\
&=e^{-\frac{i\hat{H}t'}{\hbar}}Tr_{field}\left(|\psi(t)\rangle\langle\psi(t)|\right)e^{\frac{i\hat{H}t'}{\hbar}},
\end{split}
\end{equation}
where $\hat{H}=\frac{\hat{P}^{2}}{2m}$ is the free particle Hamiltonian and $|\psi(t)\rangle$ is given by (\ref{evolution}). Thus, we can obtain the atomic distribution for a specific flight time $t'$ and observe how the initial distinguishability and visibility are tuned according to the amplitude of the quantum and classical fields. We consider that the initial atomic distribution once the atom emerges from the double-slit corresponds to two Gaussian profiles with standard deviation $\sigma=0.05\lambda_{CF}/2\pi$ and centred in the positions $x=0$ and $x=0.25\lambda_{CF}$, respectively. For each studied case, the corresponding pattern on the screen is obtained considering three different stages. First, we consider a typical double-slit scheme where we can manipulate only the parameters $c_{\uparrow}$, $c_{\downarrow}$ and $\gamma$ to define $V_{0}$, $D_{0}$ and $C_{0}$ as the initial visibility, dintinguishability and concurrence in absence of both fields. Subsequently, we add the quantum field and obtain the corresponding atomic distributions of each case. Finally, we consider the double slit with both, classical and quantum fields.

\subsection{Stage 1: Atom passing through the double slit (no fields)}
\begin{figure}[h!]\centering
\subfigure[\label{Figure8a}]{\includegraphics[width=4cm]{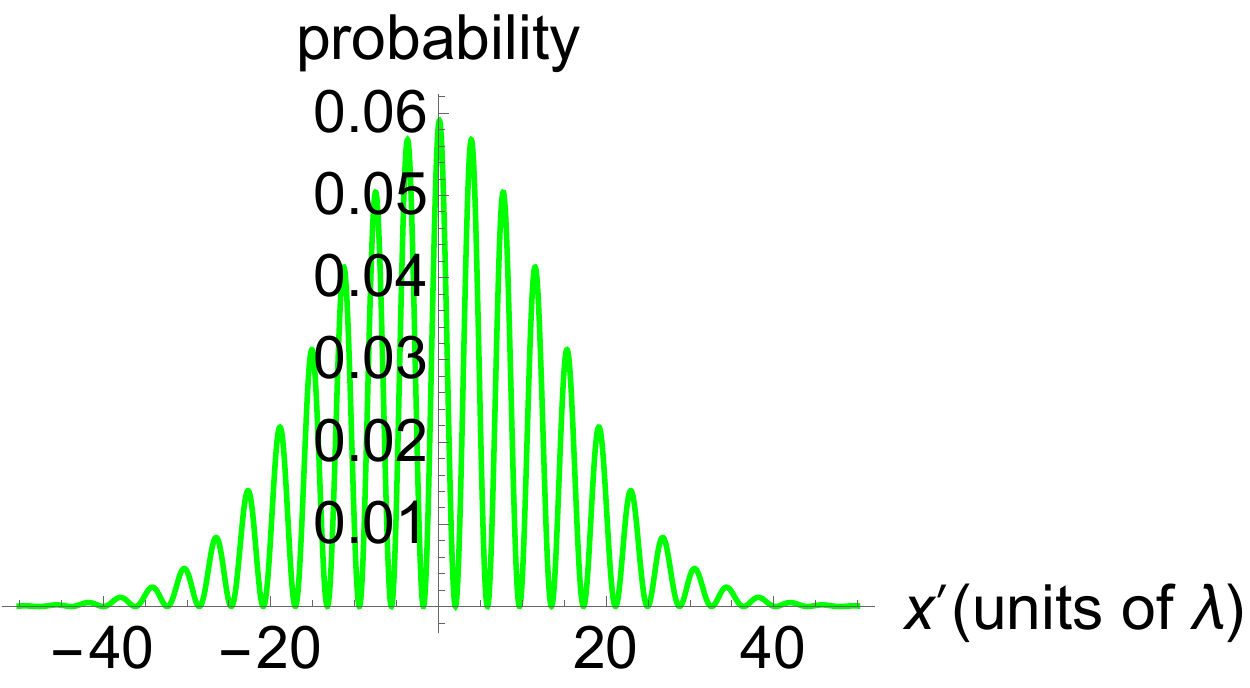}}
\subfigure[\label{Figure8b}]{\includegraphics[width=4cm]{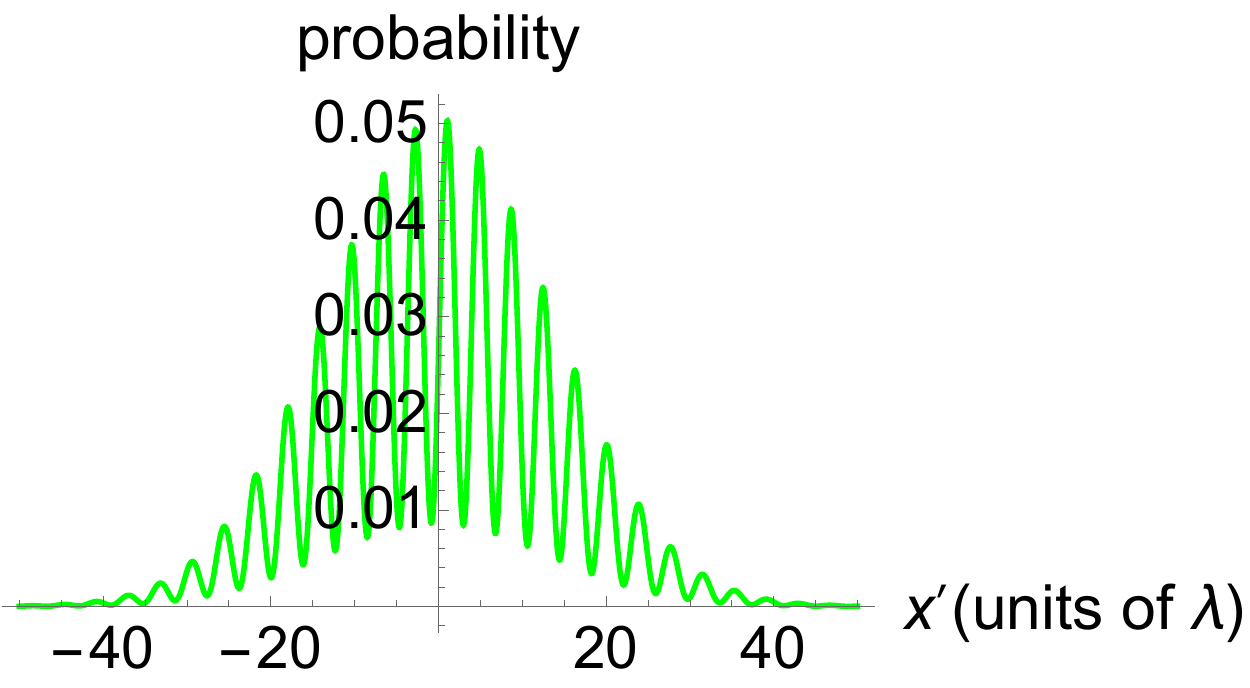}}
\subfigure[\label{Figure8c}]{\includegraphics[width=4cm]{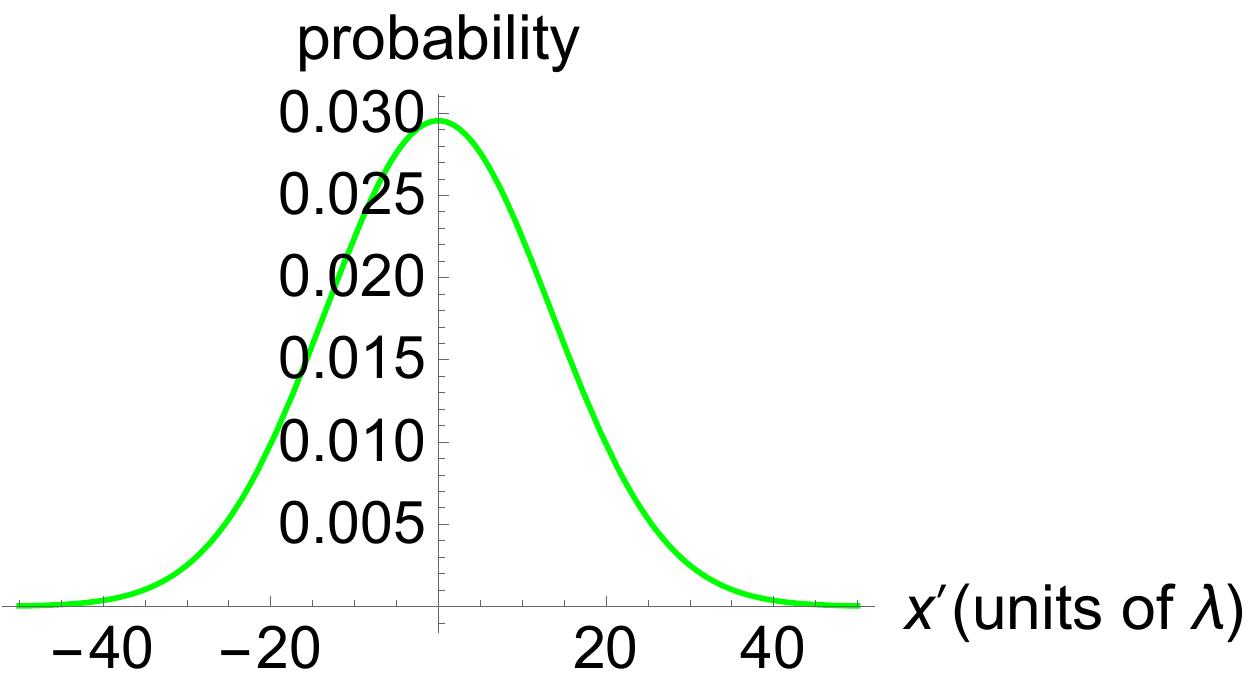}}
\subfigure[\label{Figure8d}]{\includegraphics[width=4cm]{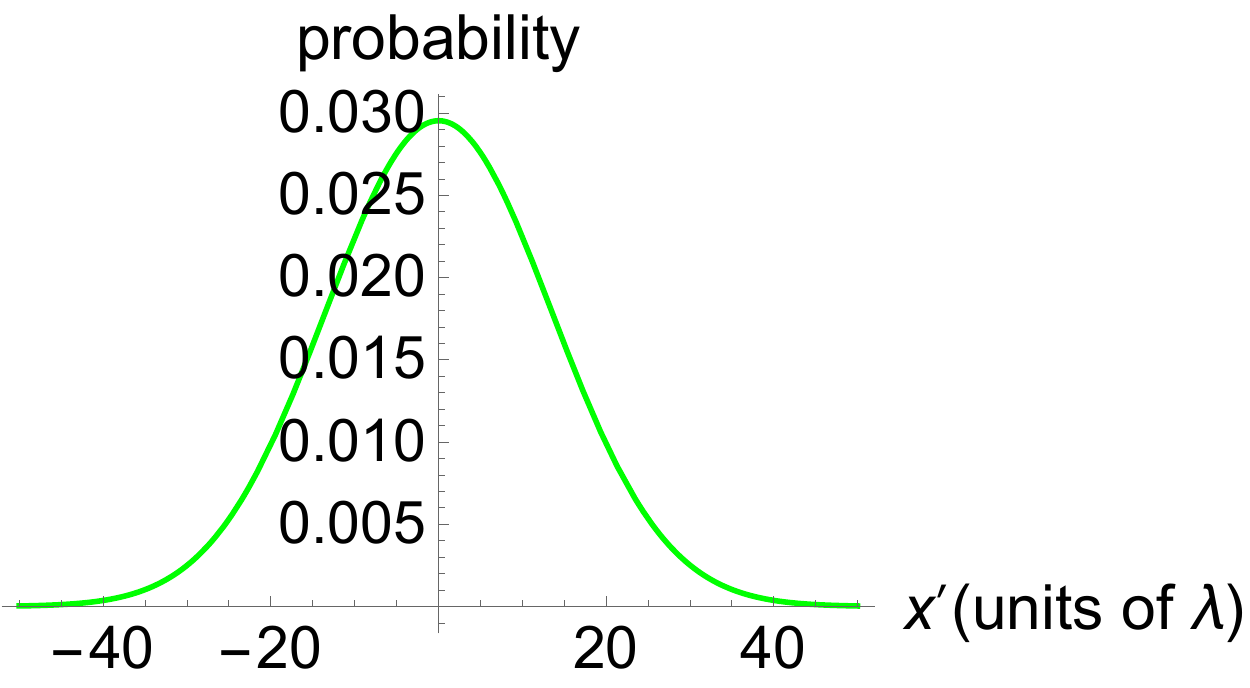}}
\subfigure[\label{Figure8e}]{\includegraphics[width=4cm]{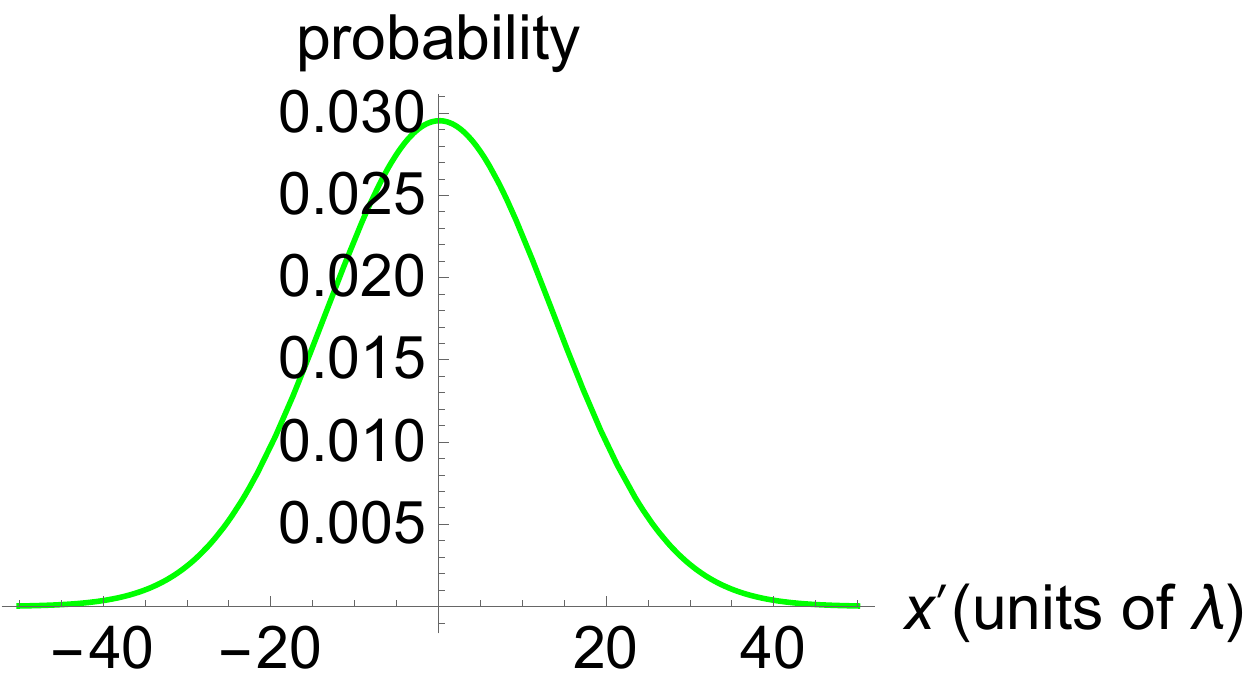}}
\subfigure[\label{Figure8f}]{\includegraphics[width=4cm]{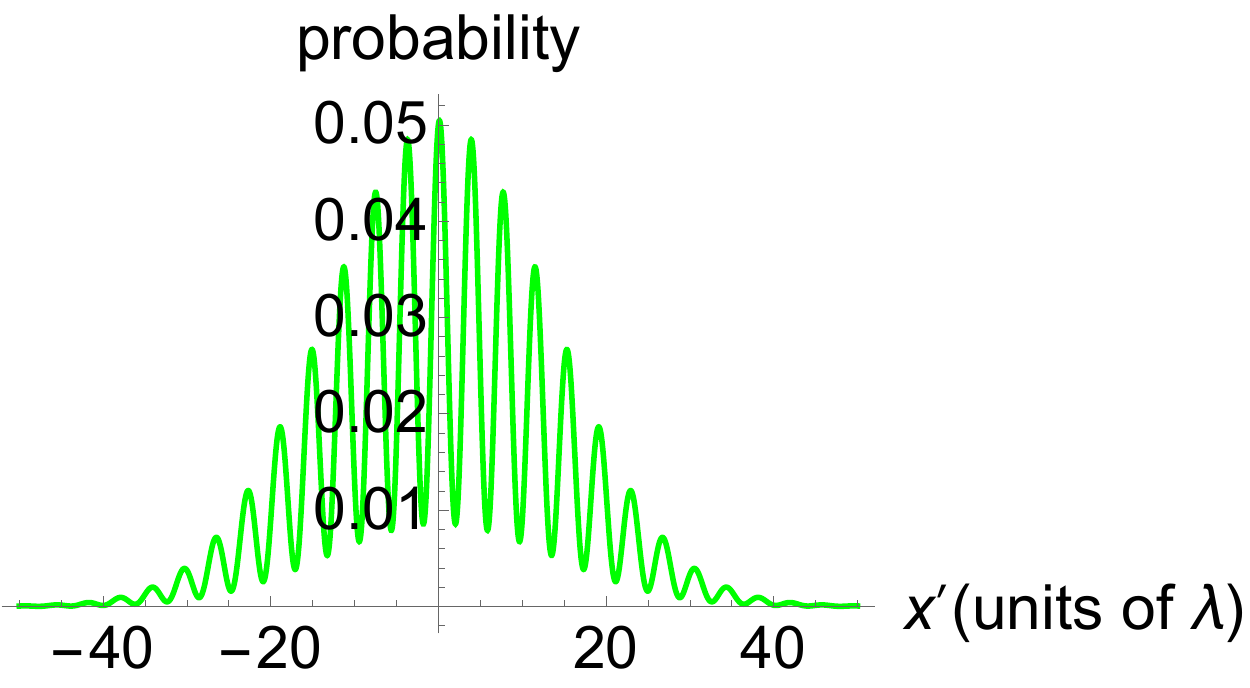}}
\subfigure[\label{Figure8g}]{\includegraphics[width=4cm]{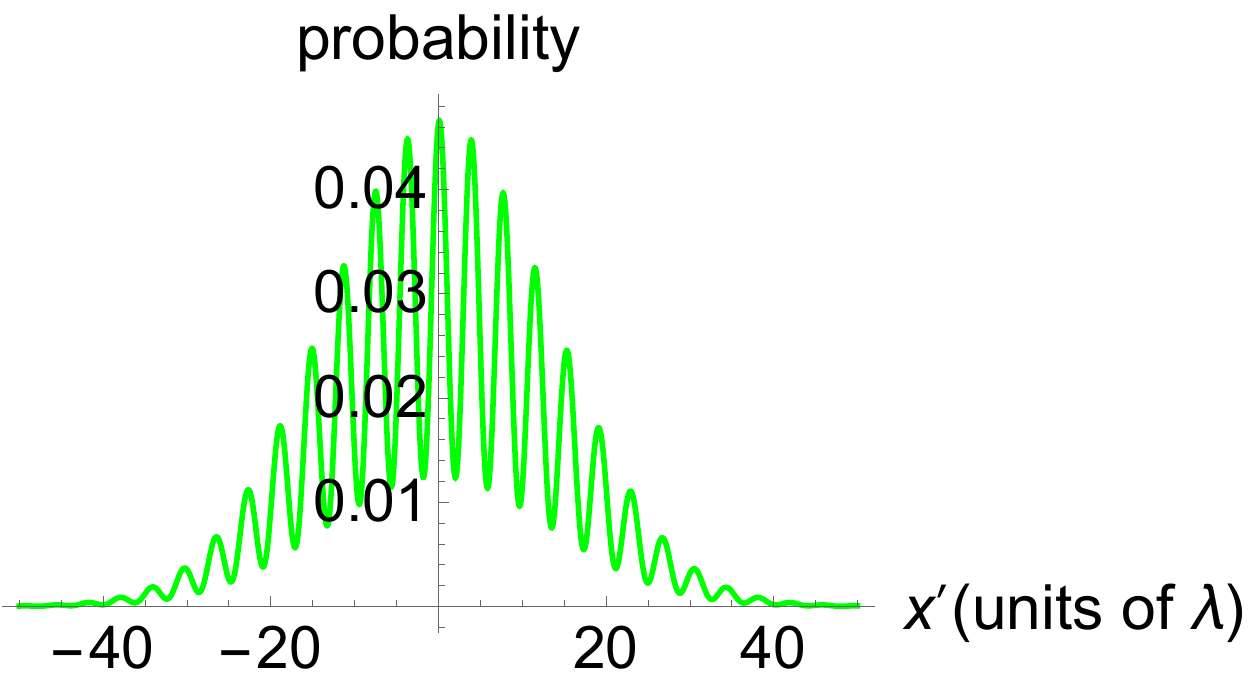}}
\caption{\underline{\textit{Stage 1}}: Atomic probability distribution obtained for each case shown on the sphere $V_{0}^{2}+D_{0}^{2}+C_{0}^{2}=1$ for $t'=3$. The distance $x'$ is expressed in units of $\lambda=\lambda_{CF}$. a) $V_{0}=1$, b) $V_{0}=D_{0}$, c) $D_{0}=1$, d) $D_{0}=C_{0}$, e) $C_{0}=1$, f) $C_{0}=V_{0}$, g) $V_{0}=D_{0}=C_{0}$.}
\end{figure}

This is the simpler stage. Distinguishability, visibility and concurrence depend only on the choice of the coefficients of reflection $c_{\uparrow}$, transmission $c_{\downarrow}$ and $\gamma$. For instance, in the case $V_{0}=1$ the internal atomic state is $|c\rangle$ in both paths, thus $\phi=0$ and $\gamma=\cos\phi=1$, which ensures $C_{0}=0$. Furthermore, the coefficients $c_ {\uparrow}$ and $c_ {\downarrow}$ are taken to be same, then $D_{0}=0$. Therefore, this corresponds to a case of total interference that is shown in green in the Fig.~\ref{Figure8a}. The values $c_ {\uparrow}=1$, $c_ {\downarrow}=0$ and $0\leq\gamma\leq 1$ correspond to other case, $D_{0}=1$, which does not show fringes of visibility [Fig.~\ref{Figure8c}]. Perhaps the most interesting case is $C_{0}=1$ [Fig.~\ref{Figure8e}], in which there is no distinguishability nor visibility. In this case, the observed pattern on the screen is similar to the typical diffraction pattern of the case $D_{0}=1$. The rest of distributions represent intermediate cases which can be obtained considering the appropriate coefficients.

\subsection{Stage 2: Atom passing through the double slit with the quantum field}
Here we consider the quantum field with an amplitude $\alpha=\sqrt{8}$, located immediately after the double-slit (see Fig.~\ref{Figure1}). As stated earlier in the Section II.C, the quantum field can store path-information in case the atom crosses the antinode in the internal state $|c\rangle$. Otherwise (state $|b\rangle$ in the upper path, or state $|c\rangle$ lower path), the phase of the quantum field remains unaffected. Thus, we have three sources of path-information: i) the choice of the coefficients $c_{\uparrow}$ and $c_{\downarrow}$, ii) the possible phase-shift of the quantum field, and iii) the internal atomic state of the atom after double-slit. \\
$\bullet$ i) As in the \textit{stage 1}, if $c_ {\uparrow}=1$ and $c_ {\downarrow}=0$, we immediately get path information.\\
$\bullet$ ii) If we choose $c_ {\uparrow}=c_ {\downarrow}$ and $\phi=0$ ($\gamma=1$), the internal atomic state in both paths is $|c\rangle$ and the path-information is recorded in the phase of the field, and can be extracted by measuring the $X$ quadrature. \\
$\bullet$ iii) Finally, for $c_ {\uparrow}=c_ {\downarrow}$ and $\phi=\pi/2$ ($\gamma=0$), the top and bottom paths are correlated to the atomic states $|b\rangle$ and $|c\rangle$, respectively. In that case the field does not store path-information. However, path-information related to the atomic states is stored and can be obtained by measuring the internal atomic state once the atom leaves the cavity.

\begin{figure}[h!]\centering
\subfigure[\label{Figure9a}]{\includegraphics[width=4cm]{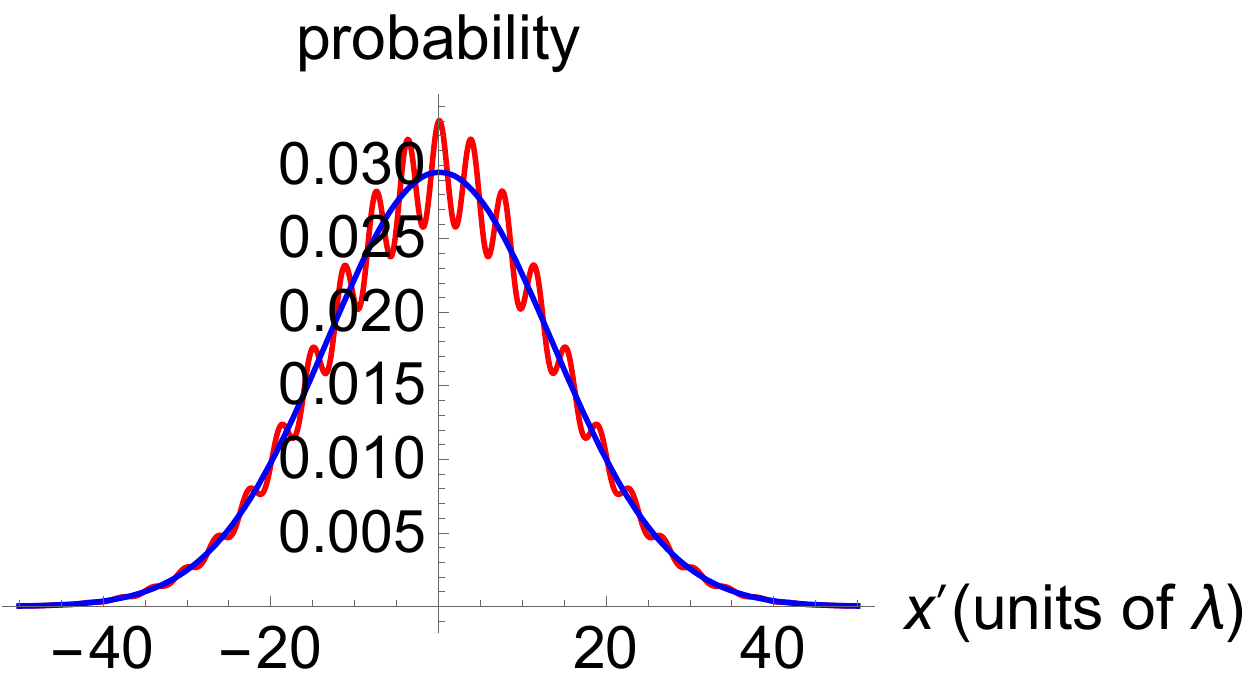}}
\subfigure[\label{Figure9b}]{\includegraphics[width=4cm]{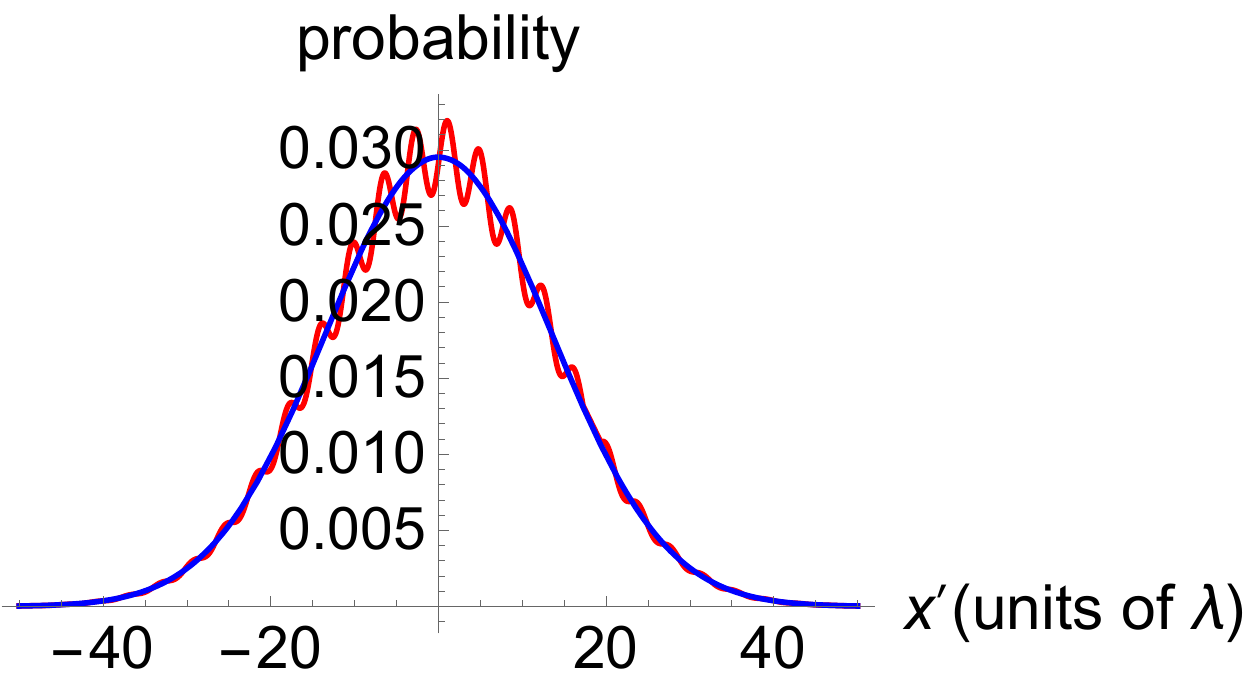}}
\subfigure[\label{Figure9c}]{\includegraphics[width=4cm]{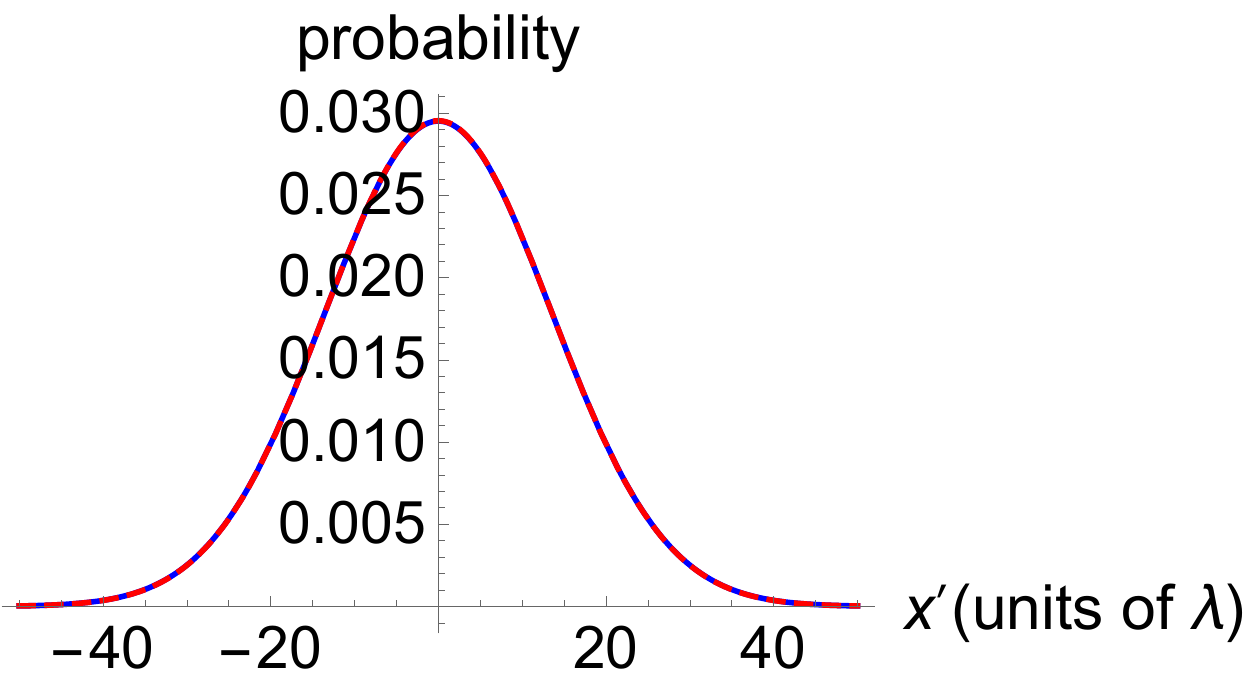}}
\subfigure[\label{Figure9d}]{\includegraphics[width=4cm]{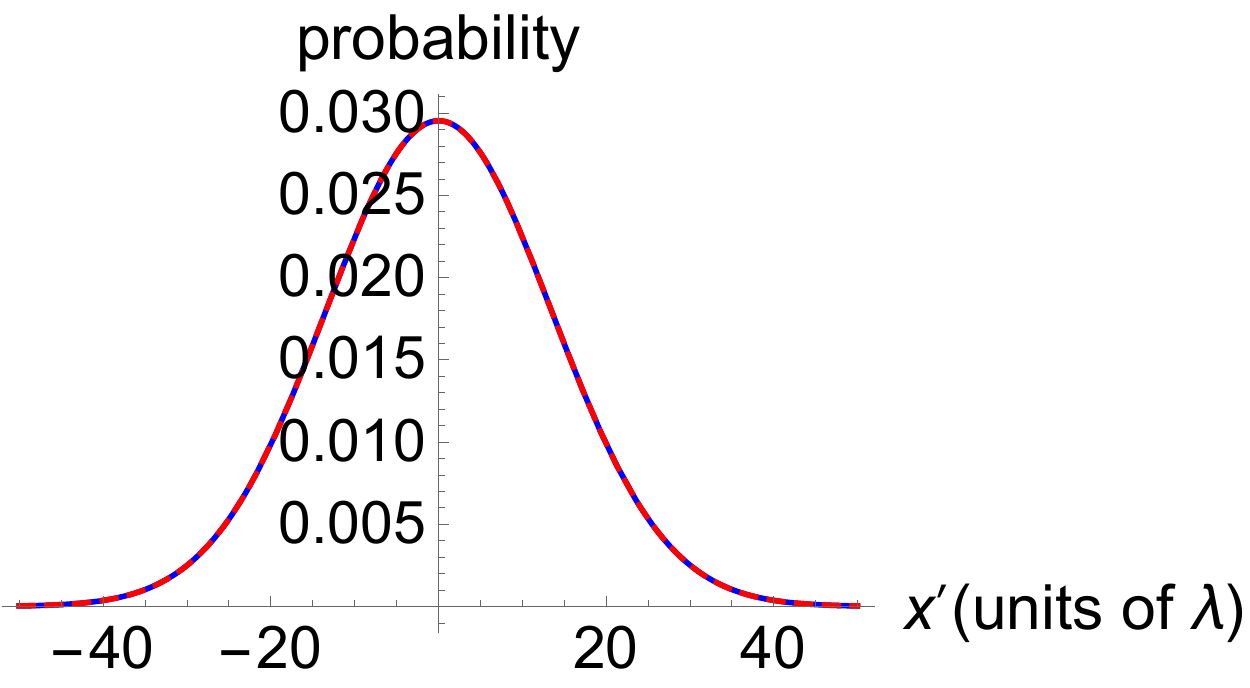}}
\subfigure[\label{Figure9e}]{\includegraphics[width=4cm]{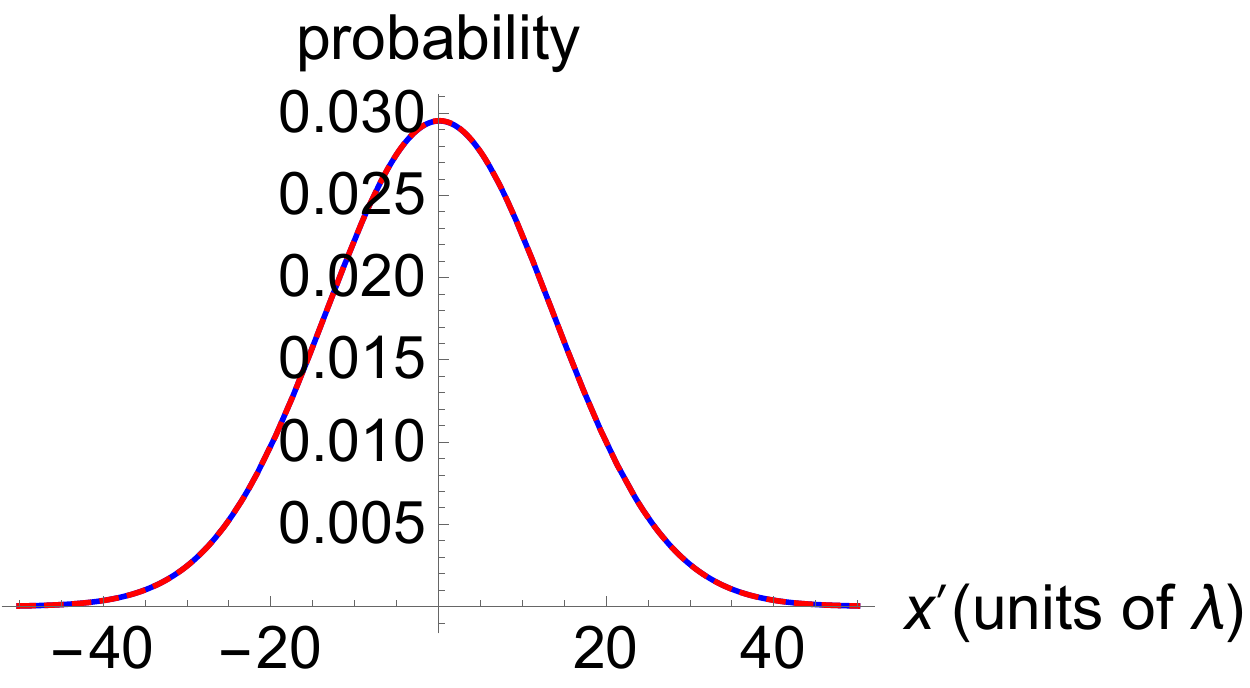}}
\subfigure[\label{Figure9f}]{\includegraphics[width=4cm]{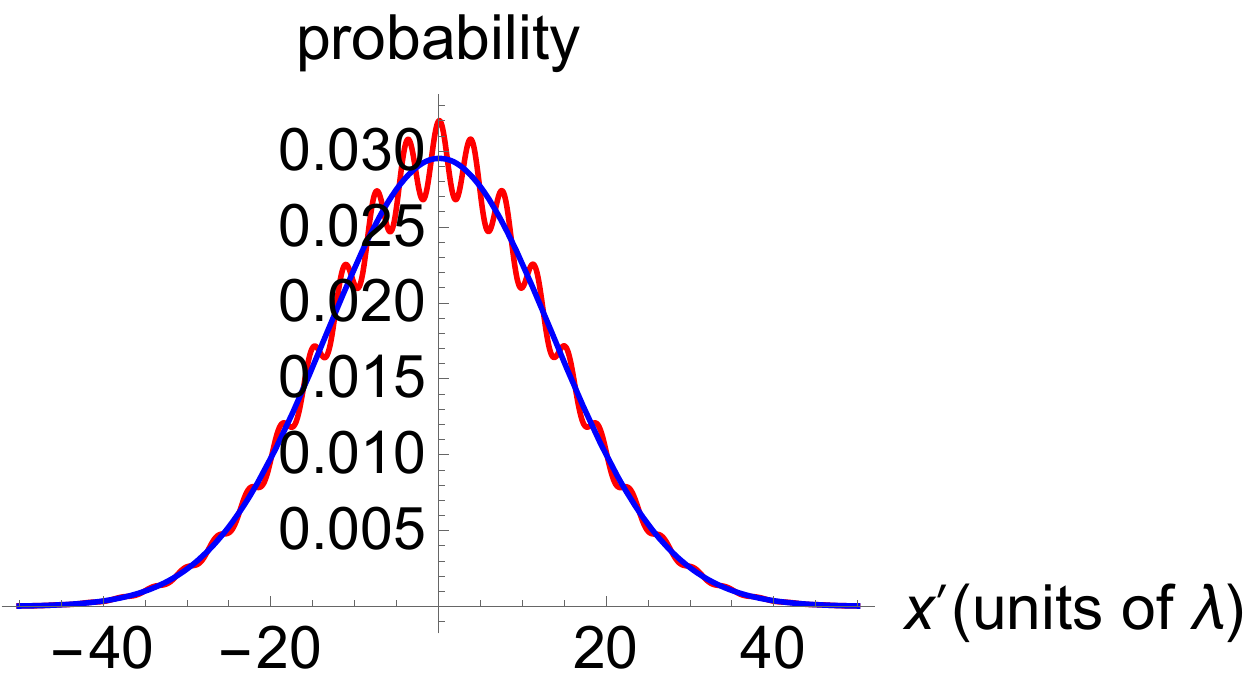}}
\subfigure[\label{Figure9g}]{\includegraphics[width=4cm]{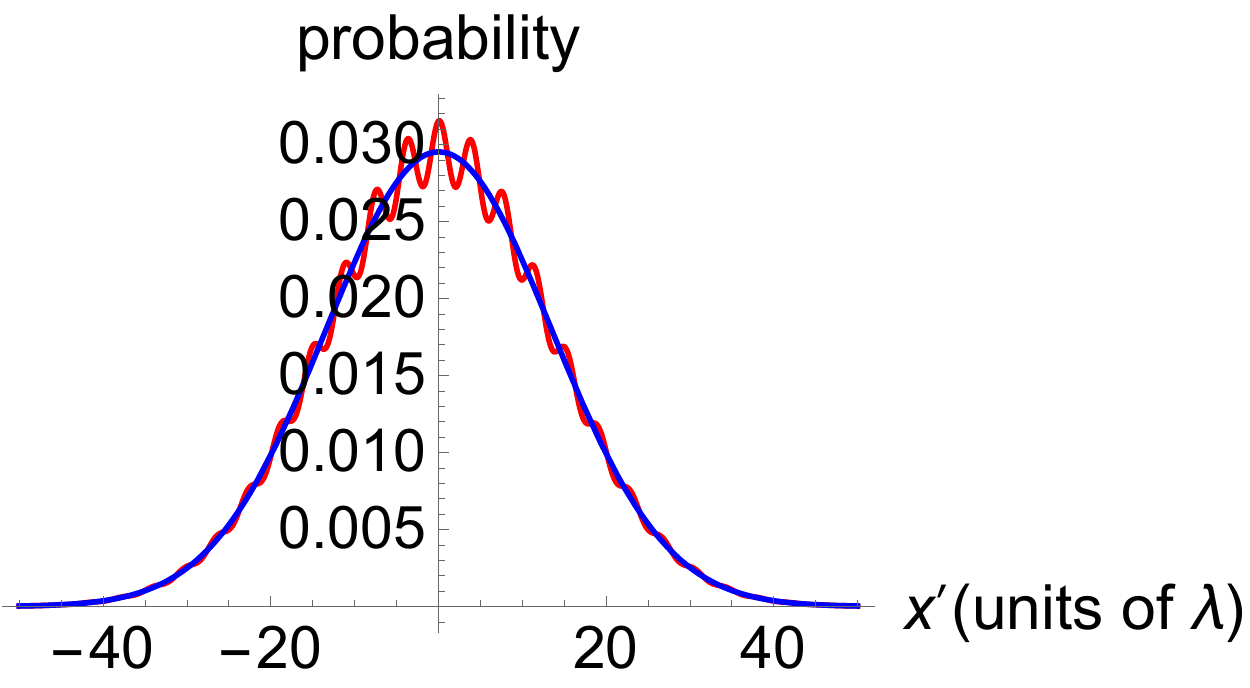}}
\caption{\underline{\textit{Stage 2}}: Atomic probability distribution obtained for each case shown on the sphere $V_{0}^{2}+D_{0}^{2}+C_{0}^{2}=1$ in presence of the quantum field for $t'=3$ with $\alpha=\sqrt{8}$ (blue) and $\alpha=1$ (red). $x'$ is expressed in units of $\lambda=\lambda_{CF}$. The choice of the parameters $c_{\uparrow,\downarrow}$ and $\gamma$ satisfies: a) $V_{0}=1$, b) $V_{0}=D_{0}$, c) $D_{0}=1$, d) $D_{0}=C_{0}$, e) $C_{0}=1$, f) $C_{0}$=$V_{0}$, g) $V_{0}=D_{0}=C_{0}$.}\label{stage2}
\end{figure}

Therefore, in presence of the quantum field we will not observe fringes of interference in any of the cases on the sphere [see blue lines in the Fig.~\ref{Figure9a} -~\ref{Figure9g}], because each case corresponds either, to one of the situations i), ii), iii), or to some intermediate state. In fact, i), ii) and iii) correspond to the cases in which the coefficients $c_{\uparrow,\downarrow}$ and $\gamma$ satisfy $D_{0}=1$, $V_{0}=1$ and $C_{0}=1$, respectively. Nevertheless, fringe visibility can be restored if the path-information is erased. In order to achieve that, the first option is reducing the amplitude of the quantum field, so that the $X$ quadrature measurement becomes ambiguous and does not reveal path-information. In this way the interference is partially restored [red lines in Fig.~\ref{Figure9a},~\ref{Figure9b},~\ref{Figure9f},~\ref{Figure9g}]. In other cases, like $D_{0}=1$ [Fig.~\ref{Figure9c}], $D_{0}=C_{0}$ [Fig.~\ref{Figure9d}] and $C_{0}=1$ [Fig.~\ref{Figure9e}], interference cannot be restored. 

A second option is performing a $Y$ quadrature measurement of the field. In this case the path-information is completely erased and interference is restored, since we assume the outcome of our measurement as the most probable result that corresponds to $\chi_{\theta=\pi/2}=0$. The green lines in the Fig.~\ref{Figure8a} -~\ref{Figure8g} are the distributions we would expect to see on the screen if a $Y$ quadrature measurement is performed on the quantum field. This is the same result that we would obtain if the quantum field were not present. 

\subsection{Stage 3:  Atom passing through the double slit with the quantum and classical fields}
Finally, we consider the double-slit scheme with both quantum and classic fields. When the classical light is present, it affects the final phase of the quantum field after the interaction, because the terms whose phases depend on $\varepsilon$ appear in the evolution operator. As a consequence, interference and path-information are altered. As in the previous stage, the phase-shift produced by $\varepsilon$ also depends on the internal atomic state $|c\rangle$ or $|b\rangle$ present in the top path.

\begin{figure}[h!]\centering
\subfigure[\label{Figure10a}$\varepsilon=1$]{\includegraphics[width=4cm]{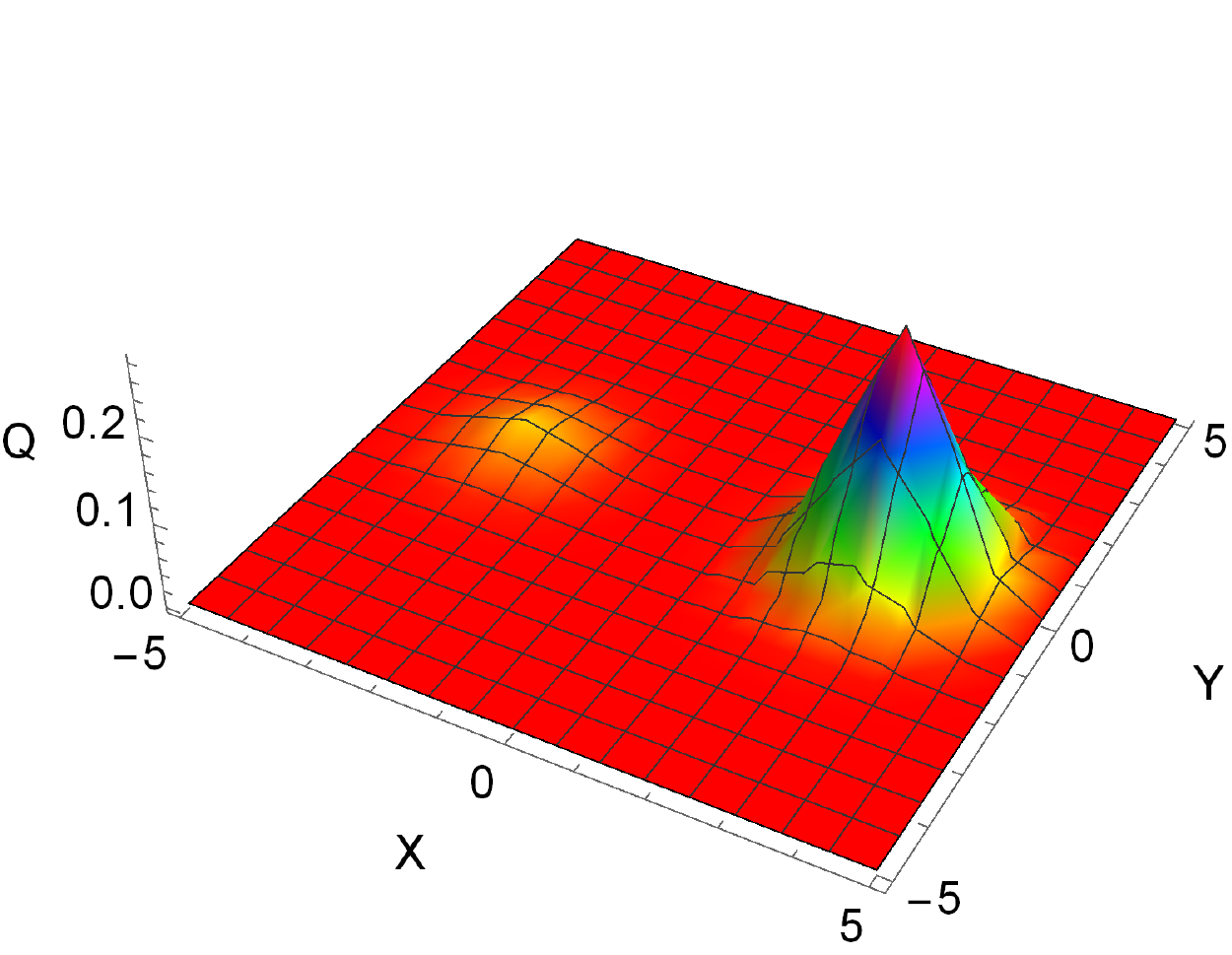}}
\subfigure[\label{Figure10b}$\varepsilon=3$]{\includegraphics[width=4cm]{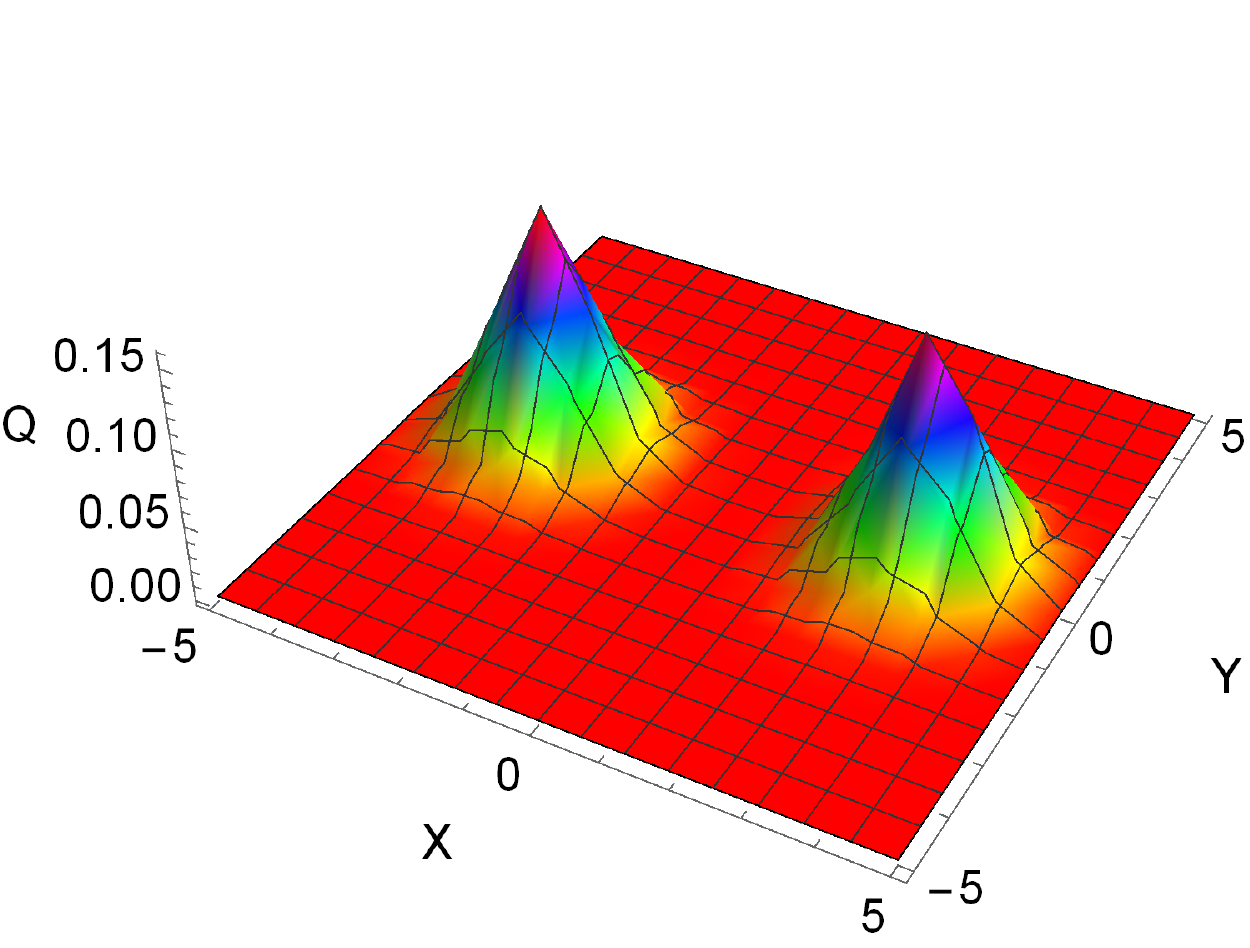}}
\subfigure[\label{Figure10c}$\varepsilon=5$]{\includegraphics[width=4cm]{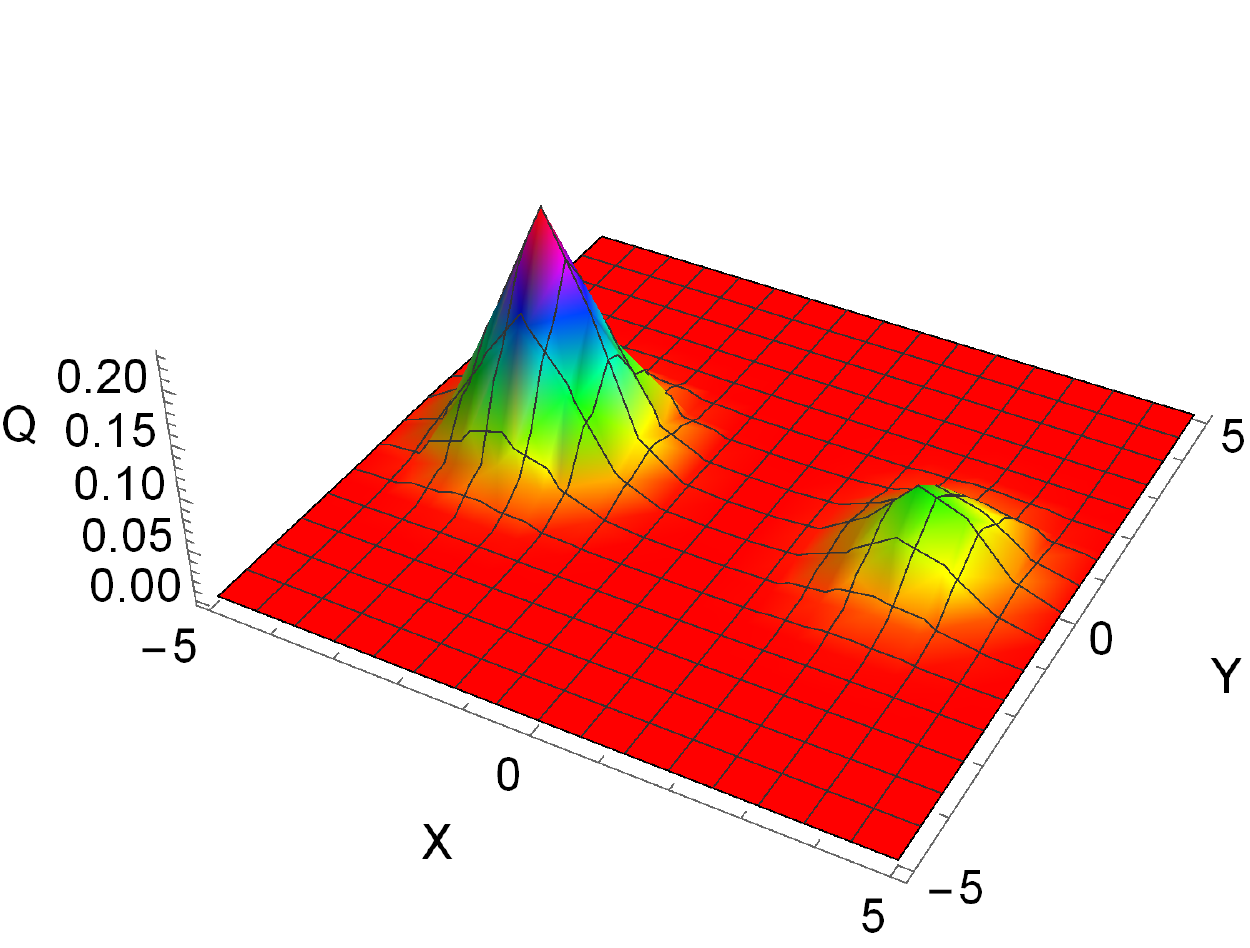}}
\subfigure[\label{Figure10d}$\varepsilon=9$]{\includegraphics[width=4cm]{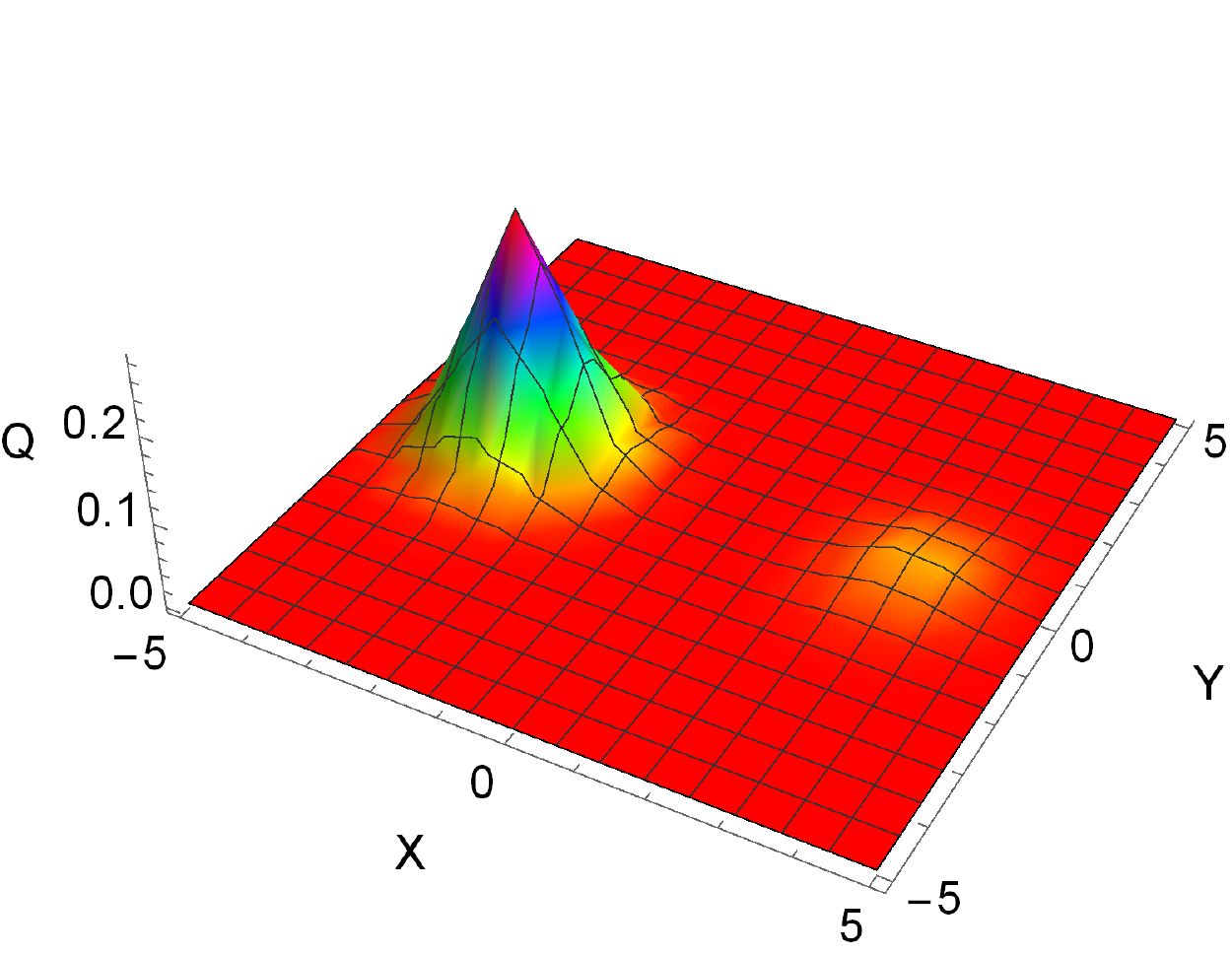}}
\caption{Internal atomic state $|b\rangle$ in the top path: As the value of $\varepsilon$ rises, the phase of the quantum field begins to differentiate from the initial phase. Thus, now a $X$ quadrature measurement can reveal path-information.} \label{Figure10}
\end{figure}

$\bullet$ \underline{\textit{The top path and internal state $|b\rangle$}}: When $\varepsilon=0$, we have already seen that the phase of the quantum field does not change and thus we cannot obtain which-path information. However, for different values of $\varepsilon$, the phase of the quantum field moves away from its initial value and then we are able to get distinguishability (Fig.~\ref{Figure10}). Therefore, the higher the value of $\varepsilon$ the more path-information we get, at the expense of visibility.

\begin{figure}[h!]\centering
\subfigure[\label{Figure11a}$\varepsilon=1$]{\includegraphics[width=4cm]{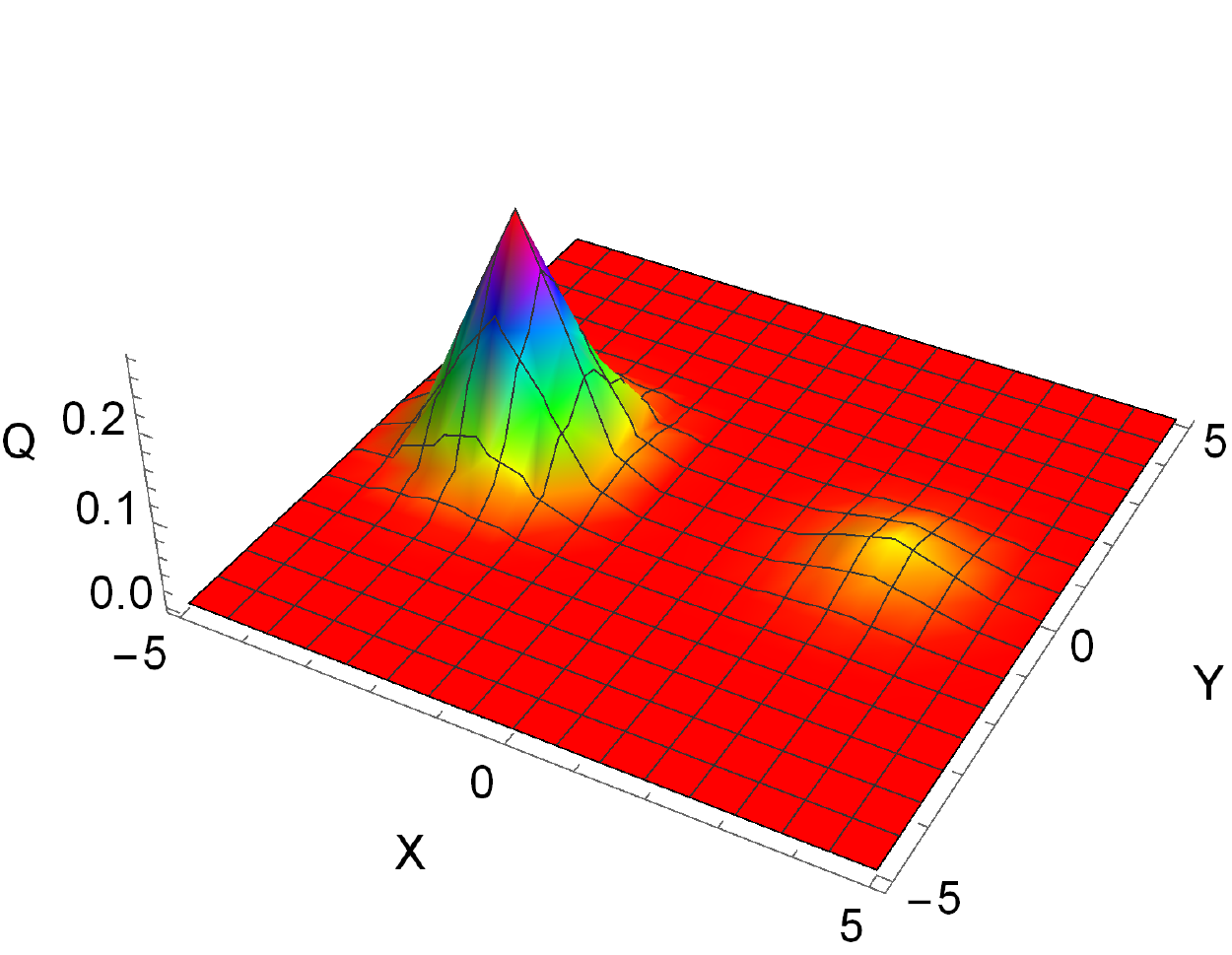}}
\subfigure[\label{Figure11b}$\varepsilon=3$]{\includegraphics[width=4cm]{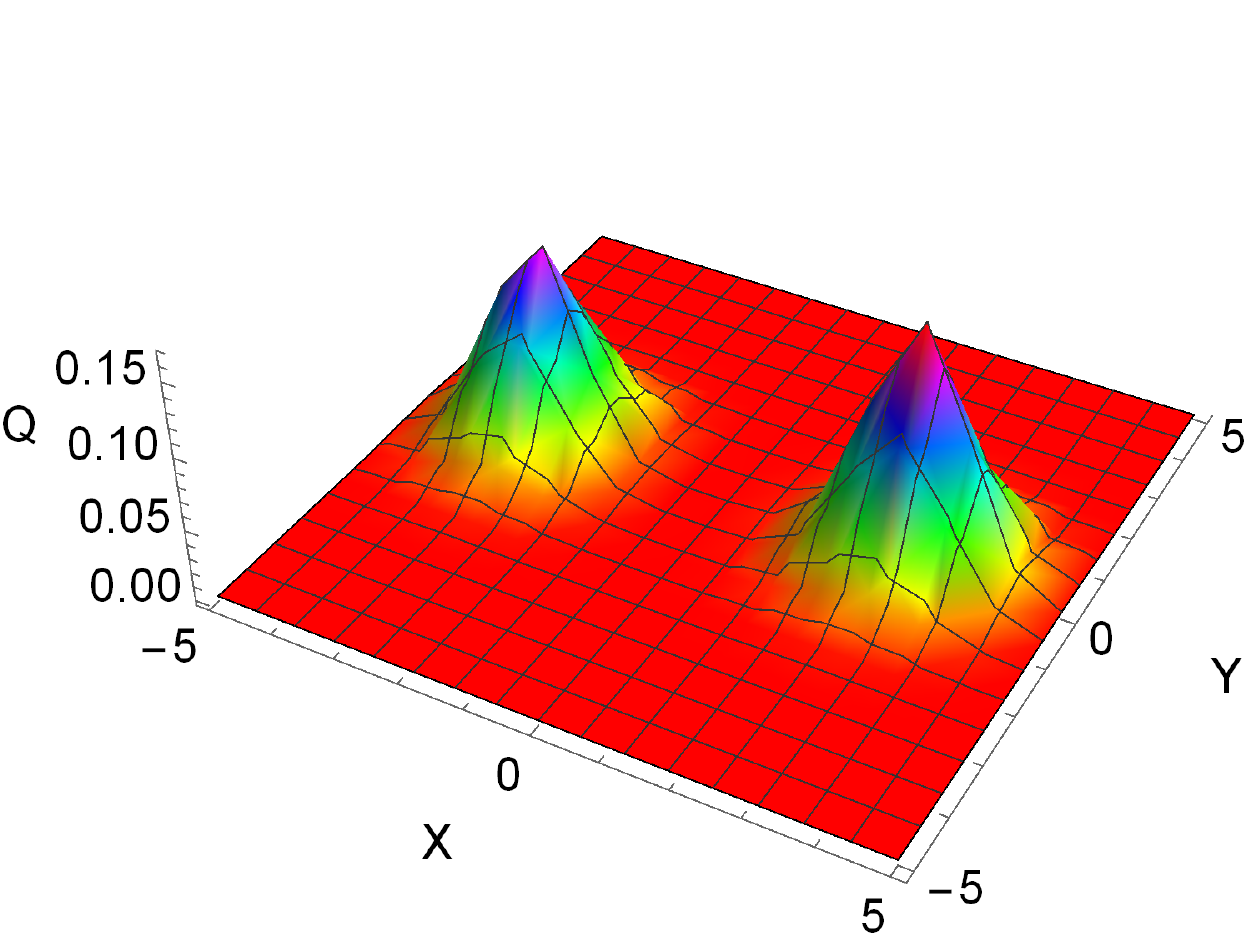}}
\subfigure[\label{Figure11c}$\varepsilon=5$]{\includegraphics[width=4cm]{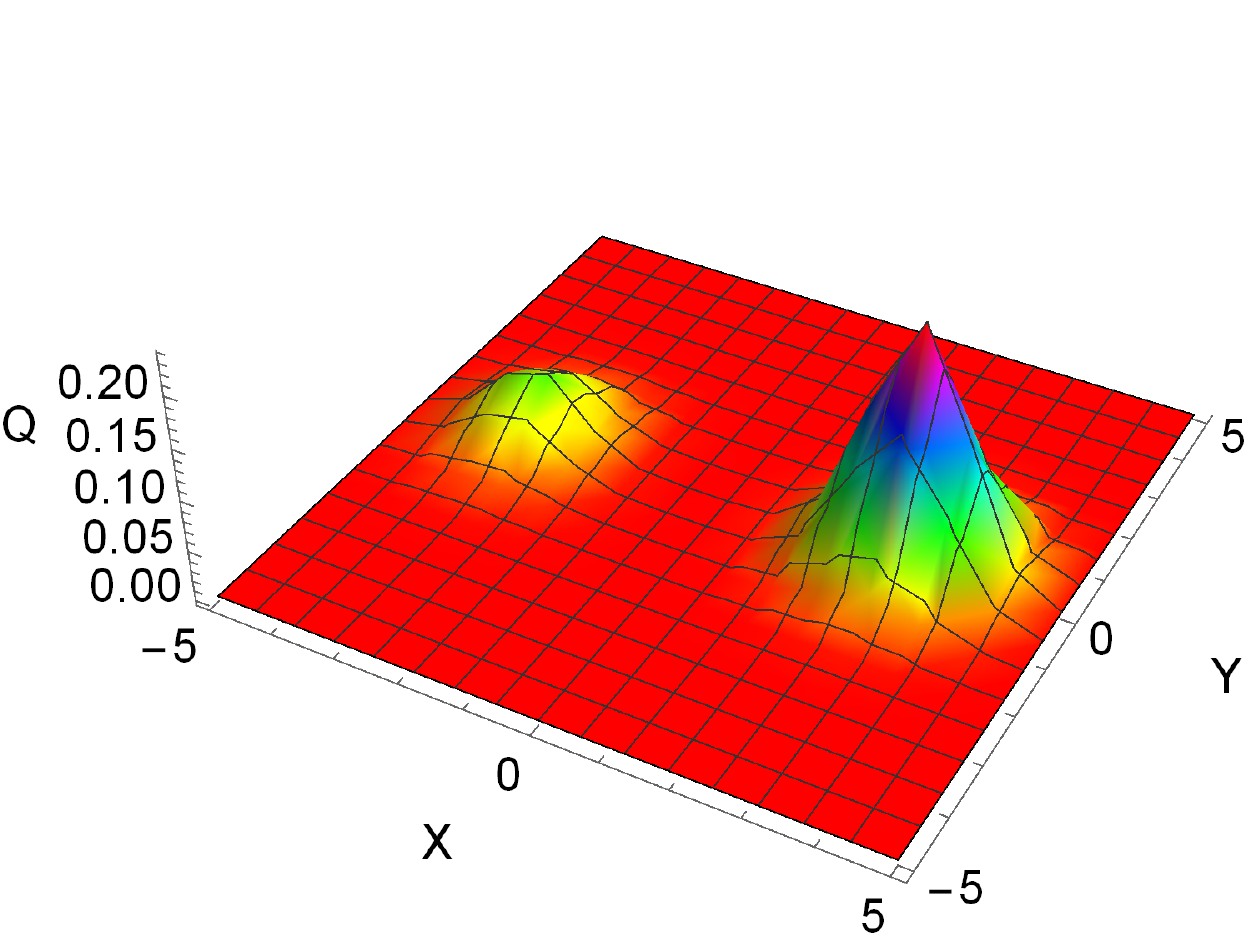}}
\subfigure[\label{Figure11d}$\varepsilon=9$]{\includegraphics[width=4cm]{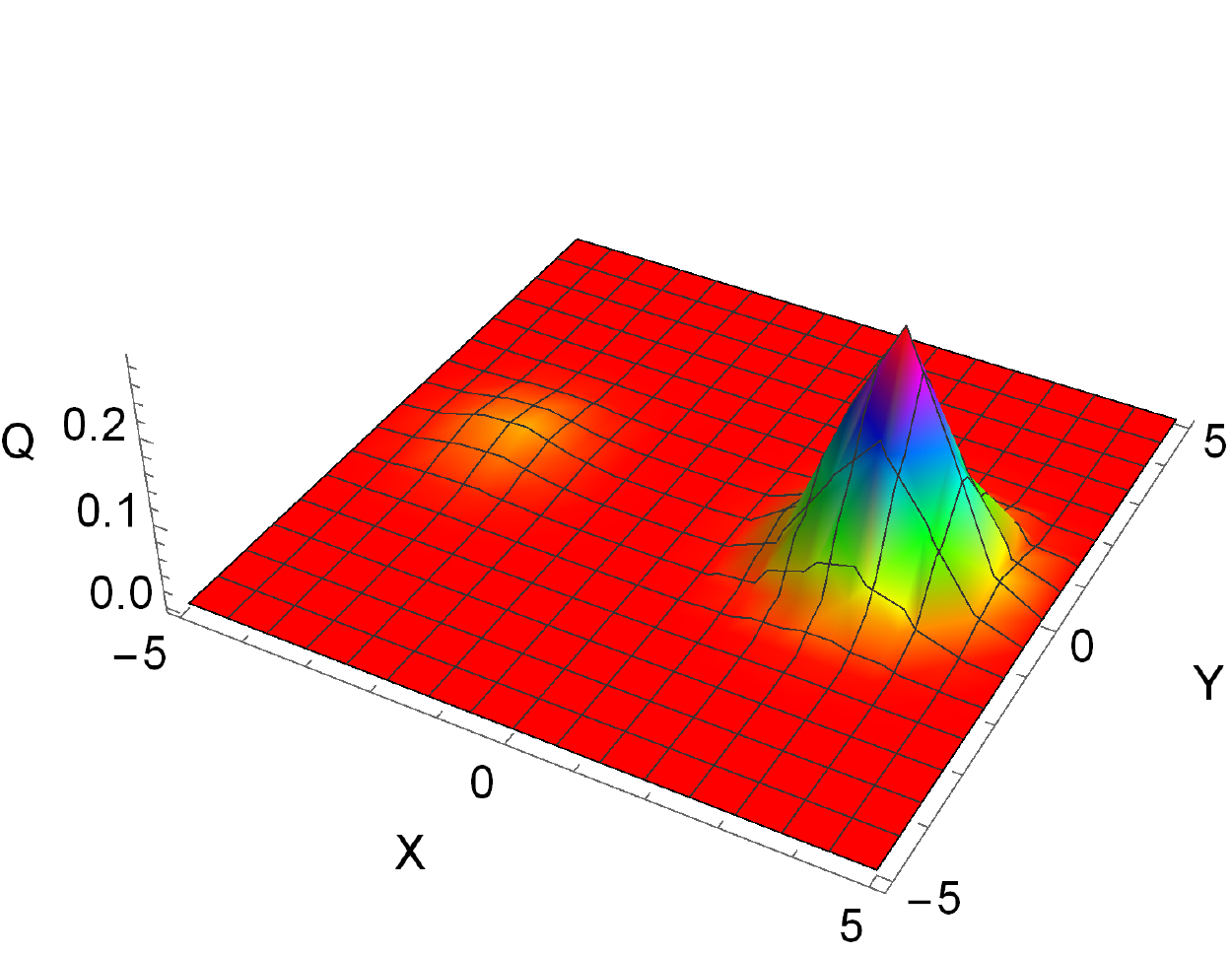}}
\caption{Internal atomic state $|c\rangle$ in the top path: In this case, as $\varepsilon$ increases, the phase of the quantum field approaches to its initial value. So now, the $X$ quadrature measurement becomes ambiguous and the path-information decreases.} \label{Figure11}
\end{figure}

$\bullet$ \underline{\textit{The top path and internal state $|c\rangle$}}: In this case, starting from $\varepsilon=0$, as we increase the classical field, the $X$ quadrature measurement becomes ambiguous, decreasing the which-path information and therefore increasing the visibility(Fig.~\ref{Figure11}).

\begin{figure}[h!]\centering
\subfigure[\label{Figure12a}]{\includegraphics[width=4cm]{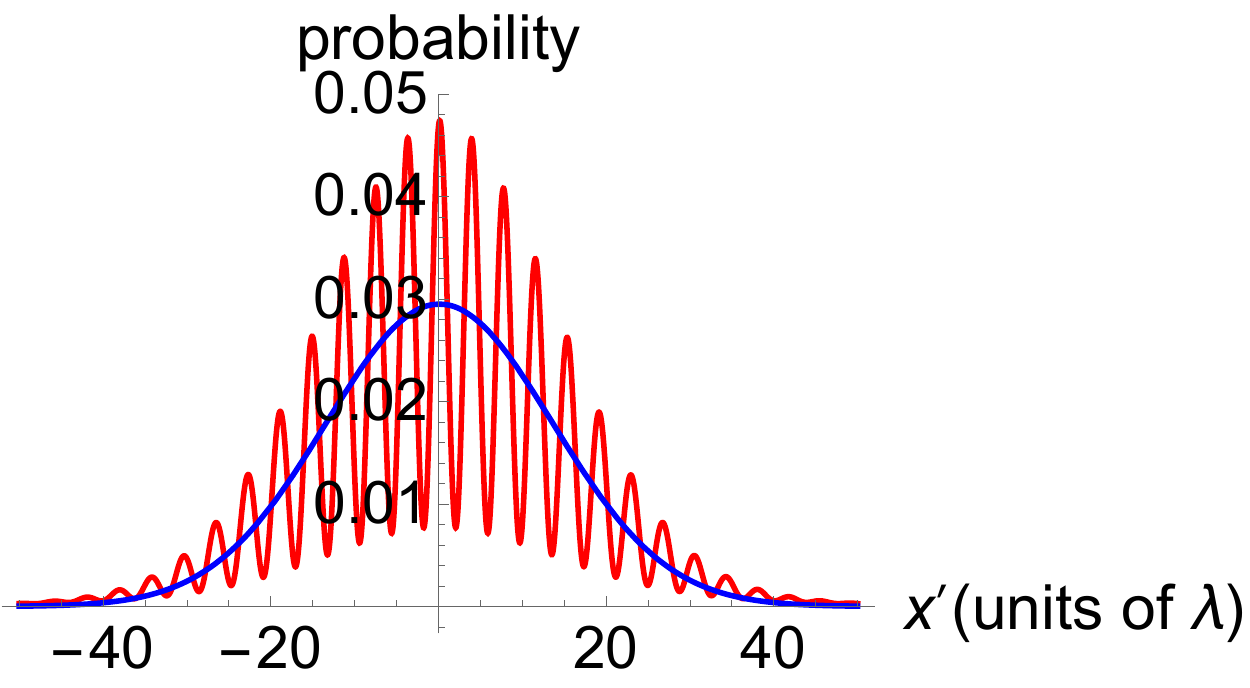}}
\subfigure[\label{Figure12b}]{\includegraphics[width=4cm]{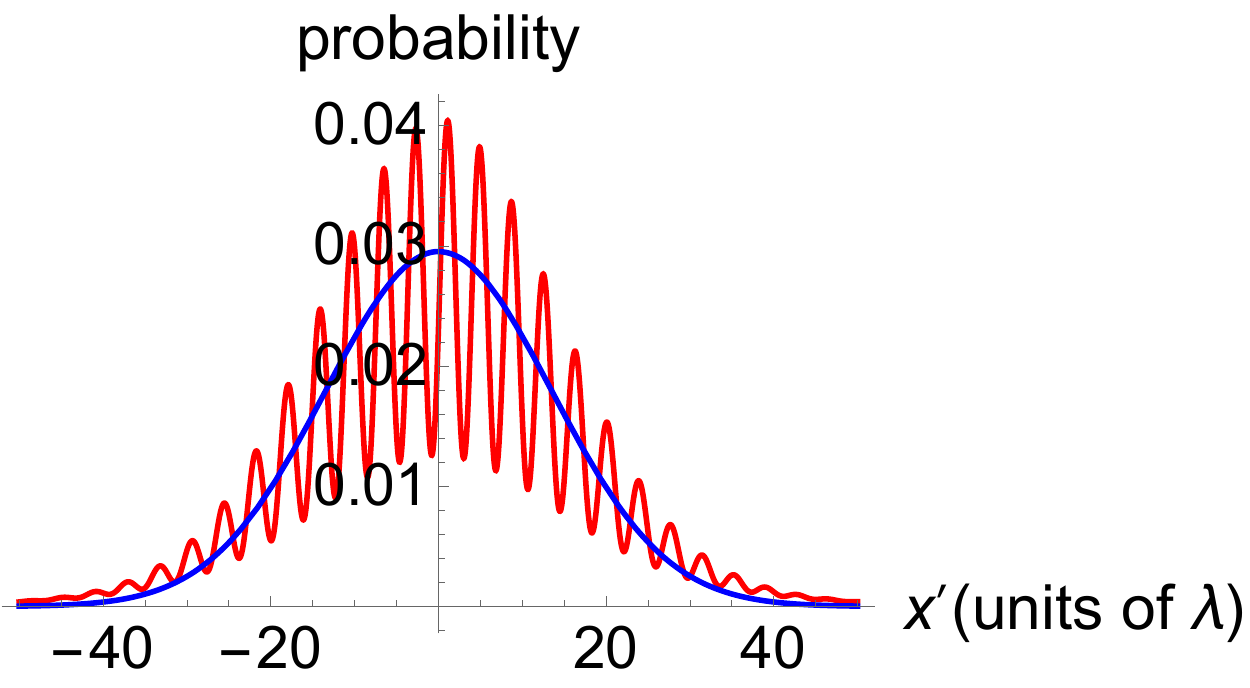}}
\subfigure[\label{Figure12c}]{\includegraphics[width=4cm]{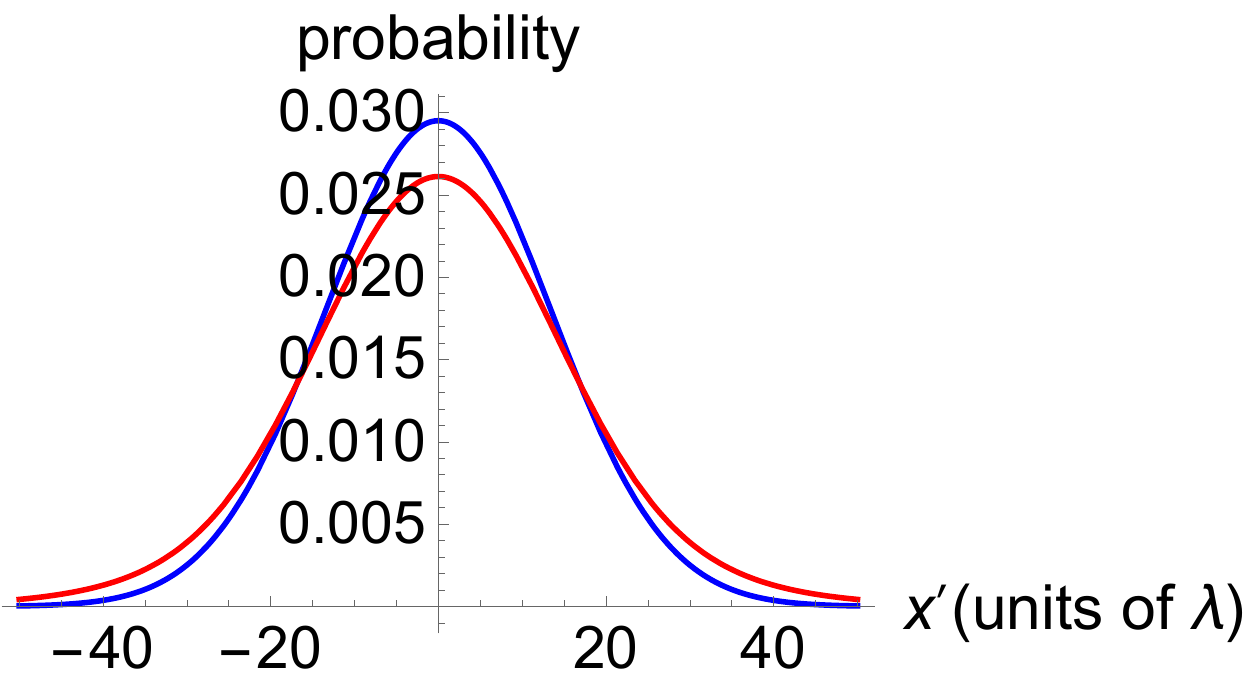}}
\subfigure[\label{Figure12d}]{\includegraphics[width=4cm]{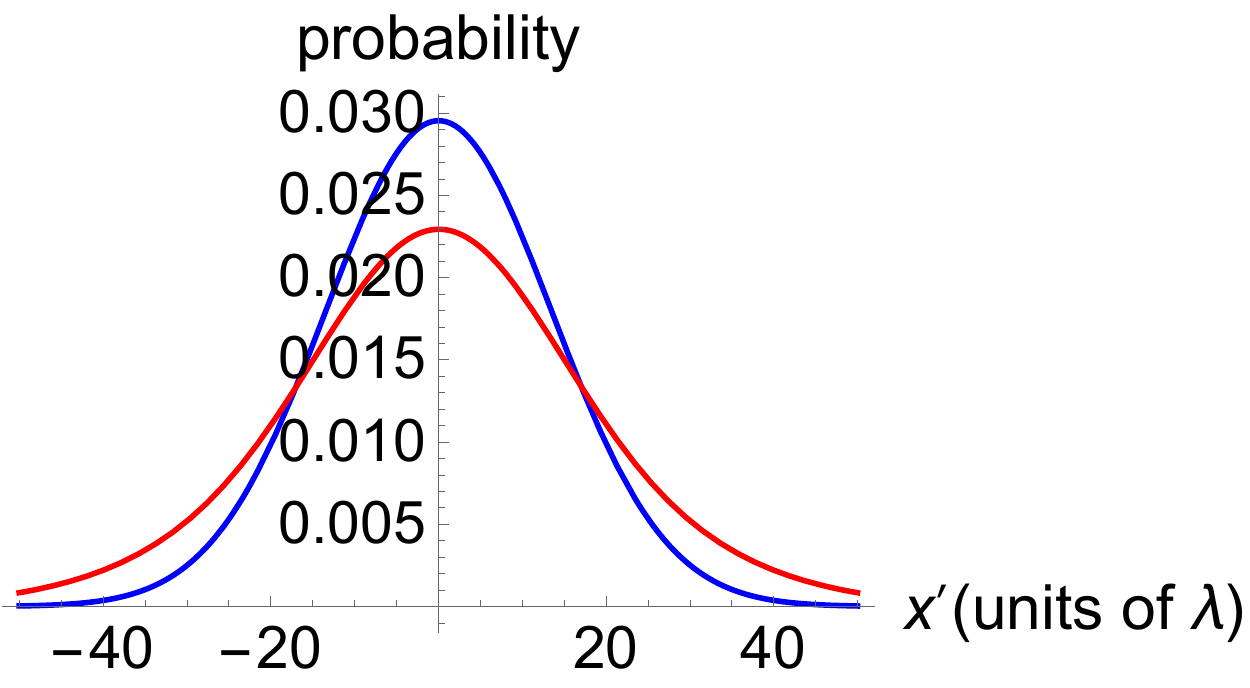}}
\subfigure[\label{Figure12e}]{\includegraphics[width=4cm]{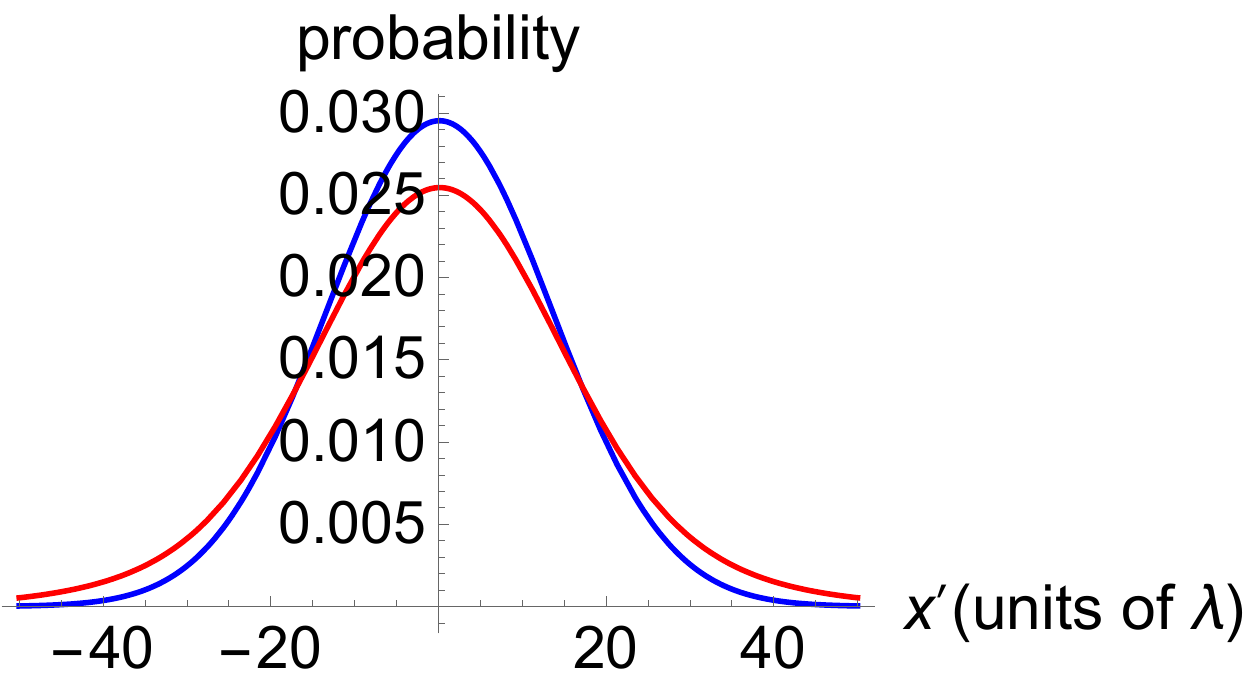}}
\subfigure[\label{Figure12f}]{\includegraphics[width=4cm]{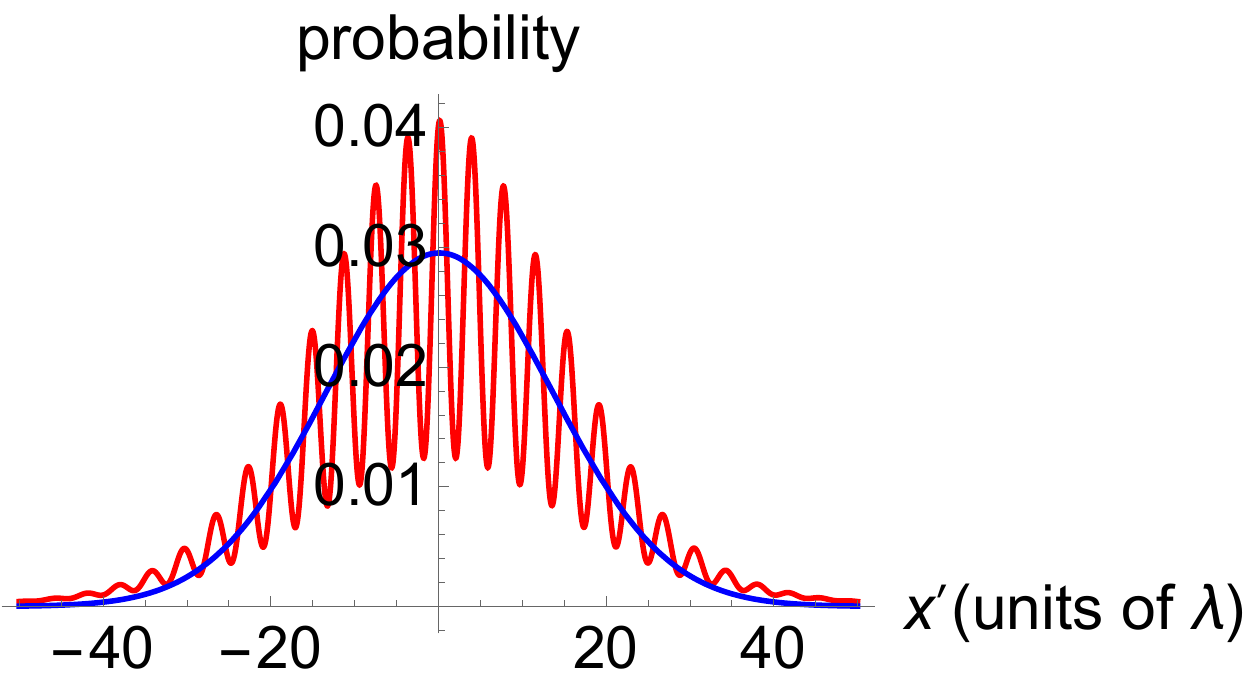}}
\subfigure[\label{Figure12g}]{\includegraphics[width=4cm]{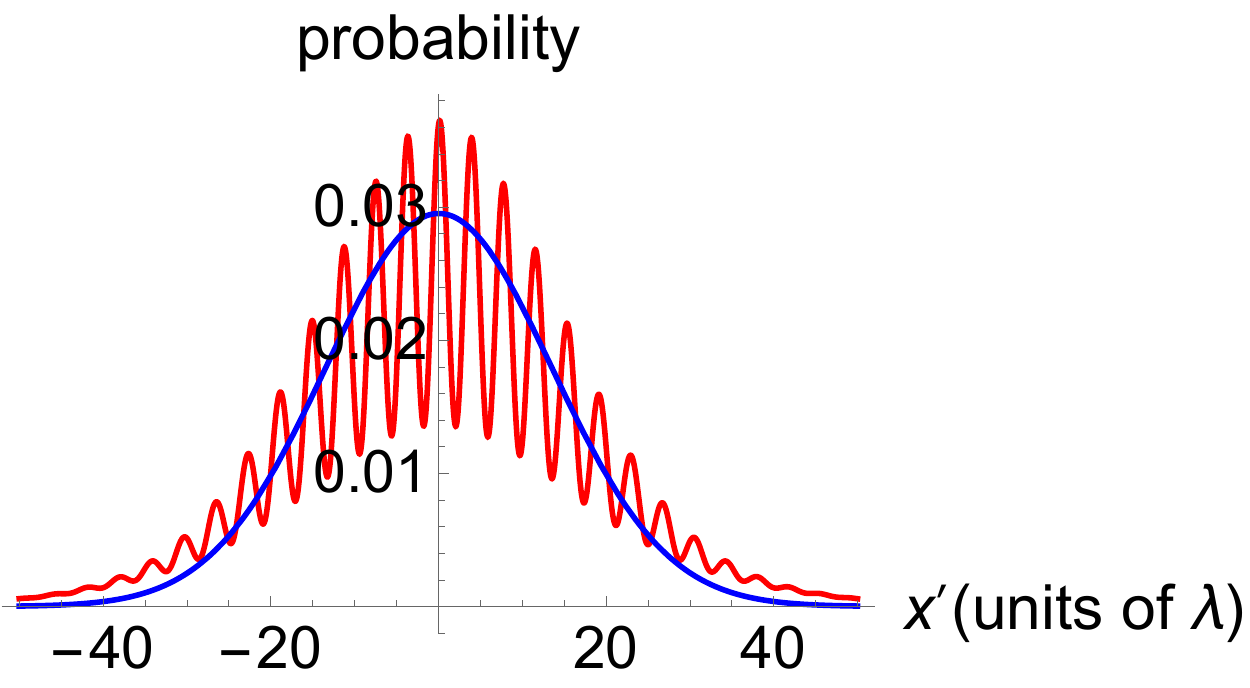}}
\caption{\underline{\textit{Stage 3}}: When $\varepsilon=3$, the effects of the atomic states $|b\rangle$ and $|c\rangle$ on the phase of the quantum field are similar [see Fig.~\ref{Figure10b} and Fig.~\ref{Figure11b}]. Therefore, a $X$ quadrature measurement cannot reveal completely path-information and the atomic distributions show partial interference in some cases and a faster evolution in other ones (red lines). Blue lines correspond to the results obtained for $\varepsilon=0$ in the \textit{stage 2}. The cases a) $V_{0}=1$, b) $V_{0}=D_{0}$, c) $D_{0}=1$, d) $D_{0}=C_{0}$, e) $C_{0}=1$, f) $C_{0}=V_{0}$, g) $V_{0}=D_{0}=C_{0}$ represent the choice of the parameters used in the \textit{stage 1}. The flight time is taken $t'=3$ with $x'$ in units of $\lambda=\lambda_{CF}$.} \label{Figure12}
\end{figure}

To show the effect of the classic field on the atomic distributions, we analyse the same cases shown before, considering $\varepsilon=3$ and $\alpha=\sqrt{8}$. In the Fig.~\ref{Figure12} we can see how the visibility fringes are restored (red lines). Thus, there is less available path-information with respect to the \textit{stage 2} (blue lines).  If we look again the case $V_{0}=1$ [Fig.~\ref{Figure12a}], we see now partial interference because now there is a probability of measuring a phase $\eta=0$ and get visibility, or $\eta=\pi$ and gain path-information. Cases like $V_{0}=D_{0}$ [Fig.~\ref{Figure12b}], $C_{0}=V_{0}$  [Fig.~\ref{Figure12f}] and $D_{0}=C_{0}=V_{0}$  [Fig.~\ref{Figure12g}] also show how the interference can be restored. On other hand, in the cases $D_{0}=1$  [Fig.~\ref{Figure12c}], $D_{0}=C_{0}$  [Fig.~\ref{Figure12d}] and $C_{0}=1$  [Fig.~\ref{Figure12e}] there is no interference, but these show that the atomic distributions evolve faster. This means that the initial Gaussian profiles of the atomic distribution in the position $x=0$ and $x=0.25\lambda_{CF}$ in $t'=0$, interact with each other earlier as compared to the case $\varepsilon=0$.

\newpage
\section{Conclusion}
The interaction between the three-level atom and both fields in a double cavity, added to a double-slit scheme, allows to study the relationship between wave-particle duality and concurrence in a more general context. In order to satisfy the equation (\ref{dvc}), and considering a Young double-slit scheme, visibility, distinguishability and concurrence can be controlled by a correct choice of the parameters involved in the definition of each one of these quantities. However, the fact of adding both fields to the scheme implies that the gain of path-information and fringe visibility also depends on the amplitude of the classical ($\varepsilon$) and quantum ($\alpha$) fields. This is because the atom-field interaction can modify the initial phase of the quantum field depending on the values of these amplitudes. The phase-shift represents path-information, which can be extracted if an adequate quadrature measurement is performed. Therefore, it is possible to obtain path-information even in the case in which the choice of the parameters $c_{\uparrow}$, $c_{\downarrow}$ and $\gamma$ satisfy $V_{0}=1$ ($D_{0}=C_{0}=0$).

In this report, we have shown how the contribution of the classical radiation alters the path-information stored in the quantum field. When the atom passes by the bottom path, the interaction is null and the initial phase remains unaffected. For $\varepsilon=0$, the maximum (minimum) path-information is obtained when the internal atomic state in the upper path is $|c\rangle$ ($|b\rangle$), due to the fact that atom-field interaction produces a $\pi$ ($0$) phase-shift. Therefore, in this case, a $X$ quadrature measurement can(not) distinguish unambiguously the path followed by the atom. However, if the internal atomic state in the upper path is $|c\rangle$, as $\varepsilon$ increases, the resulting phase-shift makes the $X$ quadrature measurement ambiguous, reducing the path-information. On the contrary, if we have the internal atomic state $|b\rangle$ in the upper path, a $X$ quadrature measurement becomes less ambiguous, giving more path-information and less visibility. Therefore, we can consider $\varepsilon$ as controlling parameter of the wave-particle duality. This is because the classical amplitude determines the transition probabilities between the internal states $|b\rangle$ and $|c\rangle$ during the atom-field interaction. For higher values of $\varepsilon$ these transitions become more probable and thus the phases of the quantum field produced by the internal atomic states are exchanged, as it is shown in the figure Fig.~\ref{Figure10}  for a transition from $|b\rangle$ to $|c\rangle$ and in the figure Fig.~\ref{Figure11} for a transition from $|c\rangle$ to $|b\rangle$. In this sense, considering the possible transitions between the internal states of the atom, we can consider the classical radiation not only as a controlling parameter of the wave-particle duality but also as a controller of a single Raman diffraction process generated by both quantum and classical fields. On other hand, if we consider the presence of both fields with a small amplitude $\varepsilon$, the transition probabilities are reduced and the atom has a larger probability of remaining in its initial internal state. In this case the process can be described as a single Bragg diffraction process. Finally, in absence of the classical contribution, only the quantum field controls the interaction and there is no a Raman nor Bragg process. 

In addition to this, and based on the different patterns observed in each case, we also conclude that for $\varepsilon$ different from zero, the atomic distributions evolve faster as compared to the $\varepsilon=0$ case. This means that a certain pattern observed on the screen in absence of the classical field, can be equally obtained but in less time if it is turned on. This is because higher values of $\varepsilon$ generate faster oscillations of the terms present in the evolution operators described in the expressions  (\ref{bb}), (\ref{cb}), (\ref{cc}) and (\ref{bc}). Therefore, the initial Gaussian profiles of the atomic distribution which emerge from the double cavity interact with each other at earlier times. In this sense, we can say that the classical field acts like a focusing device of the patterns on the screen.  
      
A curious observation. The CV plane (Fig.~\ref{Figure7}) shows that starting from $C_{0}=1$, we can recover partially or completely the interference pattern by just varying the internal atomic degrees of freedom without resorting to the distinguishability ($D_{0}=0$).    
      
Finally, an interesting case is $C_{0}=1$, in which $V_{0}$ and $D_{0}$ vanish. Our scheme shows that neither visibility nor distinguishability can be restored once the maximum concurrence has been established. Therefore, this proves the sturdiness of this case against any quadrature measurement in any of the three stages presented in the previous sections. 

\begin{acknowledgments}
We wish to acknowledge the financial support from the project FONDECYT (CL) 1180175 and Beca De Doctorado Nacional CONICYT (CL) 21171247 during the development of this research. 
\end{acknowledgments}

\appendix
\section{Effects of the evolution operator on the initial state of the quantum field $|\alpha\rangle$ in (\ref{evolution}).}\label{appa}

The elements $U_{bb}$ (\ref{bb}) and $U_{cb}$ (\ref{cb}) represent the evolution of the system when the internal atomic state is $|b\rangle$. On other hand, the elements $U_{cc}$ (\ref{cc}) and $U_{bc}$ (\ref{bc}) describe the evolution when the internal state is $|c\rangle$. If the atom crosses the lower slit ($c_{\downarrow}=1$) and then the common node in $x=0.25\lambda_{CF}=0.75\lambda_{QF}$, no interaction occurs and the quantum field remains the same (see \ref{nointeraction}).

\begin{widetext}
\begin{equation}\label{bb}
\begin{split}
\bullet U_{bb}|\alpha\rangle &=
\bigg[1+\frac{g_{2}^{2}|\varepsilon|^{2}\big[e^{-i\Delta t/2 } \big(\cos\sqrt{g_{2}^{2} |\varepsilon|^{2} + g_{1}^{2}aa^{\dag}+\Delta^{2}/4}t+\frac{i\Delta}{2}\frac{\sin\sqrt{g_{2}^{2} |\varepsilon|^{2} + g_{1}^{2}aa^{\dag}+\Delta^{2}/4}t}{\sqrt{g_{2}^{2} |\varepsilon|^{2} + g_{1}^{2}aa^{\dag}+\Delta^{2}/4}}\big)-1\big]}{g_{2}^{2} |\varepsilon|^{2} +  g_{1}^{2}aa^{\dag}}\bigg]|\alpha\rangle\\
&=\sum_{m}\bigg[1+\frac{\cos^{2}(k'x)|\varepsilon|^{2}\big[ e^{i(g^{2}\cos^{2}(k'x) |\varepsilon|^{2} + g^{2}\cos^{2}(kx)(1+m))t/\Delta}-1\big]}{\cos^{2}(k'x) |\varepsilon|^{2} +  \cos^{2}(kx)(1+m)}\bigg]e^{-\frac{|\alpha|^{2}}{2}}\frac{\alpha^{m}}{\sqrt{m!}}|m\rangle\\
&\equiv\sum_{m}\alpha_{m}^{b}|m\rangle\\
\end{split}
\end{equation}
\end{widetext}
\begin{widetext}
\begin{equation}\label{cb}
\begin{split}
\bullet U_{cb}|\alpha\rangle &=
\bigg[g_{1}g_{2}\varepsilon a^{\dag}\frac{\big[ e^{-i\Delta t/2} \big( \cos\sqrt{g_{2}^{2} |\varepsilon|^{2} + g_{1}^{2}aa^{\dag}+\Delta^{2}/4}t + \frac{i\Delta}{2}\frac{\sin\sqrt{g_{2}^{2} |\varepsilon|^{2} + g_{1}^{2}aa^{\dag}+\Delta^{2}/4}t}{\sqrt{g_{2}^{2} |\varepsilon|^{2} + g_{1}^{2}aa^{\dag}+\Delta^{2}/4}}\big) -1\big]}{g_{2}^{2} |\varepsilon|^{2} +  g_{1}^{2}aa^{\dag}}\bigg]|\alpha\rangle\\
&=\sum_{m}\bigg[\cos(kx)\cos(k'x)\varepsilon \sqrt{m+1}\frac{\big[ e^{i[g^{2}\cos^{2}(k'x) |\varepsilon|^{2} +  g^{2}\cos^{2}(kx)(1+m)]t/\Delta} -1\big]}{\cos^{2}(k'x) |\varepsilon|^{2} +  \cos^{2}(kx)(1+m)}\bigg]e^{-\frac{|\alpha|^{2}}{2}}\frac{\alpha^{m}}{\sqrt{m!}}|m+1\rangle\\
&\equiv\sum_{m}\beta_{m}^{b}|m+1\rangle\\
\end{split}
\end{equation}
\end{widetext}
\begin{widetext}
\begin{equation}\label{cc}
\begin{split}
\bullet U_{cc}|\alpha\rangle &=
\bigg[1+\frac{g_{1}^{2}a^{\dag}a \big[e^{-i\Delta t/2 } \big(\cos\sqrt{g_{2}^{2} |\varepsilon|^{2} + g_{1}^{2}a^{\dag}a+\Delta^{2}/4}t+\frac{i\Delta}{2}\frac{\sin\sqrt{g_{2}^{2} |\varepsilon|^{2} + g_{1}^{2}a^{\dag}a+\Delta^{2}/4}t}{\sqrt{g_{2}^{2} |\varepsilon|^{2} + g_{1}^{2}a^{\dag}a+\Delta^{2}/4}}\big)-1\big]}{g_{2}^{2} |\varepsilon|^{2} +  g_{1}^{2}a^{\dag}a}\bigg]|\alpha\rangle\\
&=\sum_{m}\bigg[1+\frac{\cos^{2}(kx)m\big[ e^{i(g^{2}\cos^{2}(k'x) |\varepsilon|^{2} + g^{2}\cos^{2}(kx)m)t/\Delta}-1\big]}{\cos^{2}(k'x) |\varepsilon|^{2} +  \cos^{2}(kx)m}\bigg]e^{-\frac{|\alpha|^{2}}{2}}\frac{\alpha^{m}}{\sqrt{m!}}|m\rangle\\
&\equiv\sum_{m}\alpha_{m}^{c}|m\rangle\\
\end{split}
\end{equation}
\end{widetext}
\begin{widetext}
\begin{equation}\label{bc}
\begin{split}
\bullet U_{bc}|\alpha\rangle &=
\bigg[g_{1}g_{2}\varepsilon^{*} \frac{\big[ e^{-i\Delta t/2} \big( \cos\sqrt{g_{2}^{2} |\varepsilon|^{2} + g_{1}^{2}aa^{\dag}+\Delta^{2}/4}t + \frac{i\Delta}{2}\frac{\sin\sqrt{g_{2}^{2} |\varepsilon|^{2} + g_{1}^{2}aa^{\dag}+\Delta^{2}/4}t}{\sqrt{g_{2}^{2} |\varepsilon|^{2} + g_{1}^{2}aa^{\dag}+\Delta^{2}/4}}\big) -1\big]}{g_{2}^{2} |\varepsilon|^{2} +  g_{1}^{2}aa^{\dag}}a\bigg]|\alpha\rangle\\
&=\sum_{m}\bigg[\cos(kx)\cos(k'x)\varepsilon^{*} \sqrt{m}\frac{\big[ e^{i[g^{2}\cos^{2}(k'x) |\varepsilon|^{2} +  g^{2}\cos^{2}(kx)m]t/\Delta} -1\big]}{\cos^{2}(k'x) |\varepsilon|^{2} +  \cos^{2}(kx)m}\bigg]e^{-\frac{|\alpha|^{2}}{2}}\frac{\alpha^{m}}{\sqrt{m!}}|m-1\rangle\\
&\equiv\sum_{m}\beta_{m}^{c}|m-1\rangle\\
\end{split}
\end{equation}
\end{widetext}

\begin{widetext}
\begin{equation}\label{nointeraction}
\begin{split}
&\bullet c_{\downarrow}|P_{\downarrow}\rangle \otimes \bigg[\sum_{m}\beta_{m}^{c}|m-1\rangle|b\rangle + \sum_{m}\alpha_{m}^{c}|m\rangle |c\rangle\bigg]\\
&=\int \delta (x-\lambda_{CF}/4) dx \bigg[\sum_{m}\beta_{m}^{c}|m-1\rangle|b\rangle + \sum_{m}\alpha_{m}^{c}|m\rangle |c\rangle\bigg]\\
&=\sum_{m}\bigg[\cos\left(\frac{3\pi}{2}\right)\cos\left(\frac{\pi}{2}\right)\varepsilon^{*} \sqrt{m}\frac{\big[ e^{i[g^{2}\cos^{2}\left(\frac{\pi}{2}\right) |\varepsilon|^{2} +  g^{2}\cos^{2}\left(\frac{3\pi}{2}\right)m]t/\Delta} -1\big]}{\cos^{2}\left(\frac{\pi}{2}\right) |\varepsilon|^{2} +  \cos^{2}\left(\frac{3\pi}{2}\right)m}\bigg]e^{-\frac{|\alpha|^{2}}{2}}\frac{\alpha^{m}}{\sqrt{m!}}|m-1\rangle\\
&+\sum_{m}\bigg[1+\frac{\cos^{2}\left(\frac{3\pi}{2}\right)m\big[ e^{i(g^{2}\cos^{2}\left(\frac{\pi}{2}\right) |\varepsilon|^{2} + g^{2}\cos^{2}\left(\frac{3\pi}{2}\right)m)t/\Delta}-1\big]}{\cos^{2}\left(\frac{\pi}{2}\right) |\varepsilon|^{2} +  \cos^{2}\left(\frac{3\pi}{2}\right)m}\bigg]e^{-\frac{|\alpha|^{2}}{2}}\frac{\alpha^{m}}{\sqrt{m!}}|m\rangle\\
&=\sum_{m}e^{-\frac{|\alpha|^{2}}{2}}\frac{\alpha^{m}}{\sqrt{m!}}|m\rangle\\
&=|\alpha\rangle
\end{split}
\end{equation}
\end{widetext}

\bibliography{References}
\bibliographystyle{apsrev4-2}

\end{document}